\documentclass[twocolumn,showpacs,preprintnumbers,amsmath,amssymb]{revtex4}

\usepackage{graphicx}
\usepackage{dcolumn}
\usepackage{bm}

\newcommand{\bra}[1]{\langle#1|}
\newcommand{\ket}[1]{|#1\rangle}

\newcommand{\ltsimeq}{\raisebox{-0.6ex}{$\,\stackrel
        {\raisebox{-.2ex}{$\textstyle <$}}{\sim}\,$}}

\begin{document}

\preprint{QIP07WW2\_control\_qubit}

\title{Exchange in multi-defect semiconductor clusters: assessment of `control-qubit' architectures}

\author{W. Wu}
\author{P. T.~Greenland}
\author{A. J.~Fisher}
 \email{andrew.fisher@ucl.ac.uk}
\affiliation{UCL Department of Physics and Astronomy and London Centre for Nanotechnology,\\
University College London, Gower Street, London WC1E 6BT}

\date{\today}

\begin{abstract}
We present a variational method to calculate the exchange
interactions among donor clusters in a semiconductor.  Such clusters are candidates for a so-called
control-qubit architecture for quantum information, where the effective exchange coupling between two atoms is controlled by the electronic state of a third.  We use a combination of the effective-mass approximation and the quantum defect method;  our
variational \textit{ansatz} is particularly suited to cases where an excited state of
one of the donors (control) is partially delocalised over several
different centres, forming an analogue of an extended molecular
orbital. Our method allows calculations of the "on/off"
ratios of exchange interactions in such cases. We compare exchange
interactions when the control is in the "on" and "off"
states, and find that both the magnitude and sign of the exchange interactions
may be changed. To rationalize the sign-change, we carry out a simple Green's function perturbation-theory
calculation. This simple model
qualitatively explains the sign change and illustrates its origins
both in ring-exchange processes and in the delocalization of the control electron.  We also compute probability distributions for the coupling strengths over the ensemble of clusters, and show that excitation of the control causes narrowing of the distributions along with shifts to larger magnitudes and from anti-ferromagnetic to ferromagnetic coupling.

\end{abstract}

\maketitle

\section{\label{sec:introduction}Introduction}
Proposed silicon-based quantum computer architectures
\cite{kane,koiller01} have attracted attention because of their
promise of scalability and their potential for integration with existing CMOS architectures. Localized spins in Si
are one very promising means to represent quantum information:
spin-$\frac{1}{2}$ objects are natural two-level systems, so they
may naturally be used to embody a qubit. Furthermore, in pure Si,
electron spins are associated only with defects and so are
naturally isolated.

Implementing general quantum gates requires one to produce
entangling interactions between localized spins in order to evolve
these spin states. Previous approaches to controlling these
interactions \cite{kane} required gate electrodes positioned near
to specific highly polarizable (therefore shallow) donors; such
defects are however readily ionized except at low temperatures,
and have spin-lattice relaxation times declining rapidly with
temperature because of the presence of spin-flip Raman transitions
\cite{castner63}. In addition the presence of the gates may
introduce significant additional decoherence through their
interaction with the polarizable defects, as well as posing
formidable difficulties in the accurate positioning of the defects
relative to the electrodes \cite{schofield03,jamieson05}.

An alternative scheme for controlling the interactions, which
avoids the need for electrodes, was proposed in \cite{ams}.
Quantum bits are encoded into electron spins of \textit{deep}
donors, with the advantages of lower ionization probabiliities and
much longer spin-lattice relaxation times \cite{castner63}.
Interactions are controlled, not by gates, but by local electronic
excitations \cite{ams}. Specifically, let (A,B) be two such deep
donor atoms. Their spacing should be sufficiently large that the
ground-state interaction between donor spins is small (ideally
negligible). Controlled optical excitation \cite{amsstoneham,itoh}
can promote a "control" electron from a nearby impurity C into an
excited state that is to some extent delocalized across A, B and
C. In this excited state, there is expected to be an effective
exchange interaction induced between the two qubit spins.
Qubit-qubit interactions are therefore switched on by optical
excitation and off by (stimulated) de-excitation of the control
electron.  Note that in this picture the electronic excitation is
real, not virtual, in contrast to other schemes that have been
proposed to couple quantum dots in semiconductors
\cite{piermarocchi02}; this has the advantage that the wavelength
variations arising from the inhomogeneity of the sample can be
exploited to address it on scales much smaller than the optical
wavelength \cite{ams}, while leaving the control spins unentangled
with the qubits if the operating parameters are chosen carefully
\cite{roby}.

To build a quantitative model of these processes, we need to
understand how the exchange interactions depend on the different
types of donors involved (deep or shallow), their separations, the
presence or otherwise of electronic excitations, and the extent of
delocalization of these excitations.  Exchange interactions
between pairs of hydrogenic impurities in semiconductors have been
studied for many years
\cite{herringflicker,cullis70,andres,koiller01,koiller02,wellard03,koiller04},
and we recently showed how these approaches could be easily
generalized to pairs of deep donors with an arbitrary binding
energy \cite{ourpaper1} using a simple quantum-defect approach.
The exchange among defect clusters has been much less studied,
except where the clusters strongly influence the bulk magnetic
properties, as in for example (Ga,Mn)As \cite{timm02}, although
the general importance of multi-centre interactions in determining
exchange properties has been known for many years, notably in the
context of solid $^3\mathrm{He}$ \cite{roger83}.  Their potential
significance for quantum information processing with electron
spins was pointed out more recently
\cite{mizel04,mizel04b,woodworth06}; however this work focussed on
harmonic potential wells in highly symmetric arrangements.
Deviations from the classic form of Heisenberg exchange have also
been discussed \cite{scarola05} in the context of multiple quantum
dots, and are important when the dots are very strongly
coupled or in very large magnetic fields.

We focus on a different aspect of the problem, and present a simple variational calculation designed to
capture the essential physics: the presence of deep
and shallow impurities, a range of geometries, and the possibility
of electron delocalization in the excited state.  We adopt a
reference model in which the qubits are deep donors (for example,
Bi:Si) while the control atom is a shallow donor (e.g. P:Si), and
treat it in configurations where the control is equidistant from
the qubits.  This remaining symmetry simplifies the calculations
while enabling us to study separately the effects of qubit-qubit
and control-qubit separations.  Our basic model can be thought of
as a multi-spin generalization of the well-known Heitler-London
approach \cite{heitler27,john} to the exchange interactions in the
hydrogen molecule.  Although this approximation fails at very
large separations \cite{john} because of the neglect of
correlation terms, it is very accurate over the physically
interesting range of separations up to about $12$ effective Bohr radii, beyond
which dipolar interactions anyway become dominant
\cite{ourpaper1}.  We also show that we can rationalize our main
results using simple perturbation theory.

The remainder of the paper is organized as follows.  In
\S\ref{sec:methods}, we introduce our methods and the basis
used to describe the wave functions of deep and shallow
donors. In the \S\ref{sec:results}, we present our results for a
range of qubit and control positions. In \S\ref{sec:ringexchange}
we discuss the results and show how they can be related to the
so-called kinetic exchange arising from multi-centre interactions
by using perturbation theory; then in \S\ref{sec:statistics} we show how the distributions of exchange that would be probed by bulk measurements are influenced by the properties of the defect clusters and their degree of excitation.
We do not address the optical excitation and de-excitation processes; these will be discussed in
a forthcoming paper \cite{FELIXprl}.

\section{Method of calculation}\label{sec:methods}
\subsection{Geometry and representation of the solid}
\begin{figure}
\includegraphics[width=5cm,height=4.5cm]{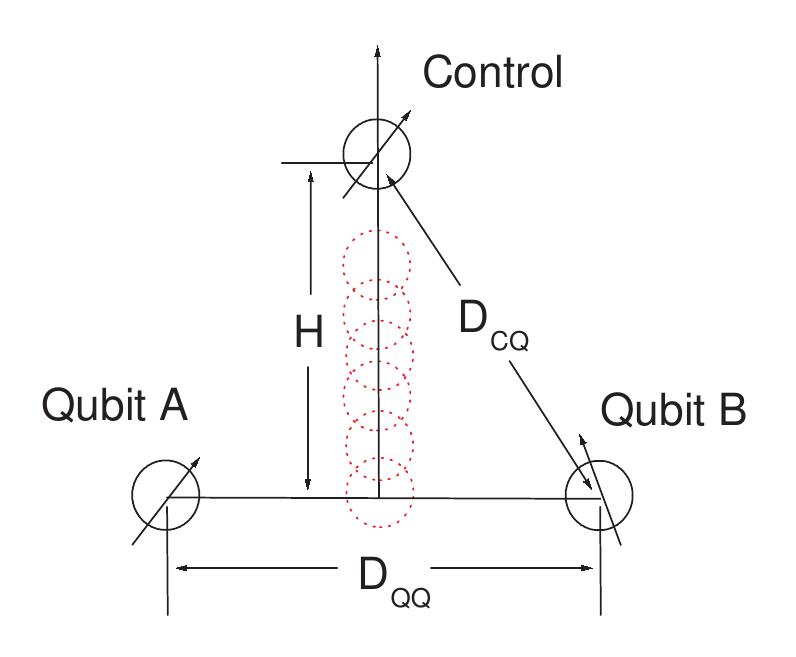}
\caption{\label{fig:geometry}The geometry used for the three-atom
system.}
\end{figure}

Let $\vec{R}_C$, $\vec{R}_A$, and $\vec{R}_B$ be the position vectors
of the control nucleus, the first qubit nucleus, and the second
qubit nucleus respectively.   We take the distance between the
qubits as $|\vec{R}_A-\vec{R}_B|=D_{QQ}$; for reasons that will become
clear below, we take the triangle formed by our three atoms to be
isosceles, so that a reflection plane exists bisecting the two
qubit atoms. Let the height of this triangle be $H$; the control
atom is then equidistant from both qubits and
$|\vec{R}_C-\vec{R}_A|=|\vec{R}_C-\vec{R}_B|=\sqrt{H^2+(D/2)^2}\equiv D_{CQ}$.
This geometry is sketched in Figure~\ref{fig:geometry}.

We treat the three donors within isotropic, single-band
effective-mass theory \cite{lk1,lk2}, introducing quantum defect
corrections \cite{bebb} where necessary to modify the binding
energies. This is the simplest model that enables us to
concentrate on the mutual interaction of the three donors, and we
have recently shown \cite{ourpaper1} that it is able to describe
the exchange between donor pairs with arbitrary binding energy.
However, it does neglect the valley degeneracy of the silicon
conduction band, which in reality contributes a
degeneracy of the donor states and corresponding interference
terms in the exchange \cite{cullis70,andres,koiller02}.  In common with other recent treatments
we do not treat the conduction-band anisotropy directly, instead
capturing its main effects by choosing an appropriate average
effective mass.

Within the effective-mass model, the Hamiltonian of this
three-electron system is
\begin{equation}
\hat{H}=\sum_{i}{\hat{h}_{i}}+\frac{1}{2}\sum_{i\neq{j}}{\frac{1}{|\vec{r}_{i}-\vec{r}_{j}|}},
\end{equation}
where
\begin{equation}
\hat{h}_{i}=-\frac{{\nabla_{i}}^{2}}{2}-\frac{Z_{C}}{|\vec{r}_{i}-\vec{R}_{C}|}-\frac{Z_{A}}{|\vec{r}_{i}
-\vec{R}_{A}|}-\frac{Z_{B}}{|\vec{r}_{i}-\vec{R}_{B}|}.
\end{equation}
Here $\vec{r}_i$ labels the coordinate of  electron $i$, and
$Z_{C}$, $Z_{A}$, and $Z_{B}$ are respectively the charges of the
control nucleus, the first nucleus, and the second nucleus. We
consider single donors and treat only the outermost electron
explicitly and therefore take $Z_C=Z_A=Z_B=1$.  Since we neglect spin-orbit
coupling and dipole-dipole interactions, the system Hamiltonian contains no explicit spin operators and the total
spin $S$ is a good quantum number.

We use scaled atomic units throughout: we take a mean
(directionally averaged) conduction-band effective mass for Si of
$m^*=0.33m_{e}$ (where $m_{e}$ is the free electron mass), and a
relative permittivity $\epsilon_r= 11.4$, leading to an effective
Bohr radius about $a_0^*=1.94\,\mathrm{nm}$ and an effective
Hartree $\mathrm{Ha}^*=33.48\,\mathrm{meV}$ \cite{faulkner69}.

\subsection{Single-centre basis functions}
We suppose there is a single important orbital $\phi_C$ at the
control site (whose nature changes depending on the degree of
excitation of the control) and two important orbitals at each
qubit site: ground-state orbitals $\chi_{i}$ (where $i\in\{A,B\}$)
which accommodate the qubit electrons, and excited orbitals
$\phi_{i}$ which allow for delocalization of the control electron.
In this paper we will assume that all of these states have $s$
symmetry, and use quantum-defect methods to represent their radial
parts.

We make this choice so that we can work equally well with deep and
shallow impurities (deep donors are especially promising for QIP
applications \cite{ams}), and still have functions that are
described by effective mass theory far from the defects.  Bebb
\cite{bebb} showed how quantum defect methods could be used
to obtain good approximate wave functions in the region outside
the impurity ion core, using only a knowledge of the energy
eigenvalues.  In this region, where the potential approaches
$u(r)\sim-1/r$, the radial Schr\"{o}dinger equation is
\begin{equation}\label{eq:asymptoticradial}
(\frac{1}{2}\frac{d^2}{dr^2}-\frac{l(l+1)}{2r^2}-u(r)-\frac{1}{2\nu^2})P(r)=0,
\end{equation}
where $-1/2\nu^2$ is just $\epsilon$, the observed energy level.
It is conventional to write $\nu=n-\mu(\epsilon_{n})$, where $n$
is the usual principal quantum number, $\mu$ is the quantum defect, and
\begin{eqnarray}
\epsilon_n=\frac{\mathrm{Ha}^{\star}}{2\nu^2}=\frac{\mathrm{Ha}^{\star}}{2(n-\mu(\epsilon_n))^2}.
\end{eqnarray}where $\mathrm{Ha}^*$ is the effective Hartree.

The solution of (\ref{eq:asymptoticradial}) that is regular at
infinity is  $P_{\nu,l}(r)$, but if $\mu\ne0$ this solution
diverges at the origin if the potential remains $1/r$ down to
short distances. We must therefore suppose that the short-range
deviation of the potential from the Coulomb form is such that the
true solution is indeed regular at the origin; the solution $P_{\nu,l}(r)$ is only valid outside the core.
In principle one could determine the entire
eigenfunction using some form of central cell correction
\cite{nara66,nara67,kerridge}.  However for our purposes the
asymptotic form is sufficient to calculate the exchange
interactions when the donor separations are large.

The normalized quantum defect wavefunction is
\begin{equation}
P_{\nu,l}(r)=N_{\nu,l}W_{\nu,l+\frac{1}{2}}(2r/\nu).
\end{equation}
Here $W$ is a Whittaker function \cite{abramovitz}, defined in
terms of hypergeometric functions $U$ by
\begin{equation}
W_{\kappa,\mu}(z)=e^{-z/2}z^{1/2+\mu}U(\frac{1}{2}-\kappa+\mu,1+2\mu,z).
\end{equation}
The approximate normalization constant $N(\nu,l)$ is given by
\cite{bebb}.

In this paper we take $\mu=0$ (and hence $\nu=1,2,\ldots$) for shallow (control) donors, and $\mu=0.3$ (hence $\nu=0.7,1.7\ldots$) for the more deeply bound qubit states.

\subsection{Variational method}\label{sec:variational}
We make a simple variational assumption for the wave function of
the control electron, writing the corresponding one-electron state
as
\begin{equation}
\psi_{C}=\alpha\phi_{C}+\beta\phi_{A}+\gamma\phi_{B}.
\end{equation}
Although this is a very simple choice, and will not reflect the
full complexity of the problem, it does give the control electron
variational freedom to adjust between the extremes of being
localized entirely on the control atom, or entirely on a
"molecular" excited state of the qubit atoms $A$ and $B$.

We then construct a three-electron variational function in the
form
\begin{widetext}
\begin{eqnarray}
\Psi_{total}^Q(x_{1},x_{2},x_{3})&=&\hat{A}[\psi_{C}(\vec{r_{1}})
(\chi_{A}(\vec{r}_{2})\chi_{B}(\vec{r}_{3})-\chi_{A}
(\vec{r}_{3})\chi_{B}(\vec{r}_{2})){Q}]
\\\Psi_{total}^{D_1}(x_{1},x_{2},x_{3})&=&\hat{A}[\psi_{C}(\vec{r_{1}})
(\chi_{A}(\vec{r}_{2})\chi_{B}(\vec{r}_{3})-\chi_{A}
(\vec{r}_{3})\chi_{B}(\vec{r}_{2})){D}_1]
\\\Psi_{total}^{D_0}(x_{1},x_{2},x_{3})&=&\hat{A}[\psi_{C}(\vec{r_{1}})
(\chi_{A}(\vec{r}_{2})\chi_{B}(\vec{r}_{3})+\chi_{A}
(\vec{r}_{3})\chi_{B}(\vec{r}_{2})){D}_0],
\end{eqnarray}
\end{widetext}
where \textbf{$Q$} refers to the quartet (total spin $S=3/2$),
\textbf{$D_1$} refers the doublet (total spin $S=1/2$) constructed
from the triplet state of the qubits, and \textbf{$D_0$} refers to
the doublet (total spin $S=1/2$) constructed from the singlet
state of the qubits. $\hat{A}$ is the antisymmetrizing operator.
The existence of a plane of symmetry in our assumed geometry means
that the functions $\Psi_{total}^{D_1}$ and $\Psi_{total}^{D_0}$,
which have opposite parities with respect to the exchange of the
qubits, cannot mix and can be varied independently.

Note that, in constructing the above function, we have neglected
high-energy `ionic' configurations in which both qubit electrons
reside on the same atom.  We have also neglected correlations
between the control electron and the qubit electrons.

We now perform a variational calculation using undetermined
multiplier methods to minimize the expectation values of the system
Hamiltonian:
\begin{eqnarray}
&&\frac{\partial{E_{i}}}{\partial{\alpha}}+\lambda\frac{\partial{S_{i}}}{\partial{\alpha}}=0\\
&&\frac{\partial{E_{i}}}{\partial{\beta}}+\lambda\frac{\partial{S_{i}}}{\partial{\beta}}=0\\
&&\frac{\partial{E_{i}}}{\partial{\gamma}}+\lambda\frac{\partial{S_{i}}}{\partial{\gamma}}=0,\\
\end{eqnarray}
where $i\in\{Q,D_1,D_0\}$, $\lambda$ is the undetermined
multiplier, $E_i=\bra{\Psi_i}\hat{H}\ket{\Psi_i}$ and
$S_i=\langle\Psi_i\ket{\Psi_i}$.

\subsection{Extraction of the spin Hamiltonian}
Once we have the energies $(E_Q,E_{D_1},E_{D_0})$, of the spin
states, we can calculate the exchange constants.  Provided spin-orbit coupling is negligible, the
three-electron spin Hamiltonian in zero external field must take
the form
\begin{equation}\label{eq:exchangeH}
\hat{H}_{eff}=J_{CQA}\vec{s}_{A}\cdot\vec{s}_{C}+J_{CQB}\vec{s}_{B}\cdot\vec{s}_{C}
+J_{QQ}\vec{s}_{A}\cdot\vec{s}_{B},
\end{equation}
because this is the only form which is (i) invariant under
simultaneous rotation of all the spins and (ii) even under
time-reversal.  Furthermore, atoms $A$ and $B$ are equivalent in
our calculation, so $J_{CQA}=J_{CQB}=J_{CQ}$. We can then
calculate the exchange constants in the effective Hamiltonian from
the eigenvalues and eigenstates.  As explained in
\S\ref{sec:variational} and Appendix~\ref{app:extractH}, the
doublet states are distinguished by the `intermediate' angular
momentum of the qubit spins (A and B); once these states have been
identified, the parameters in equation~(\ref{eq:exchangeH}) can be
extracted to give
\begin{eqnarray}\label{eq:equationsforcouplings}
J_{QQ}&=&-\frac{4}{3}E_{D_0}\nonumber\\
J_{CQ}&=&2\left(E_Q-\frac{J_{QQ}}{4}\right)
\end{eqnarray}
where the zero
of energy is chosen to be the weighted mean of the spin energies,
so the spin Hamiltonian is traceless.  Note that, with the sign
convention adopted in setting up the spin Hamiltonian
(\ref{eq:exchangeH}), positive couplings $J$ correspond to
antiferromagnetic interactions.

\section{Variational calculation: results}\label{sec:results}
\begin{figure}
\begin{tabular}{c}
\includegraphics[height=5cm]{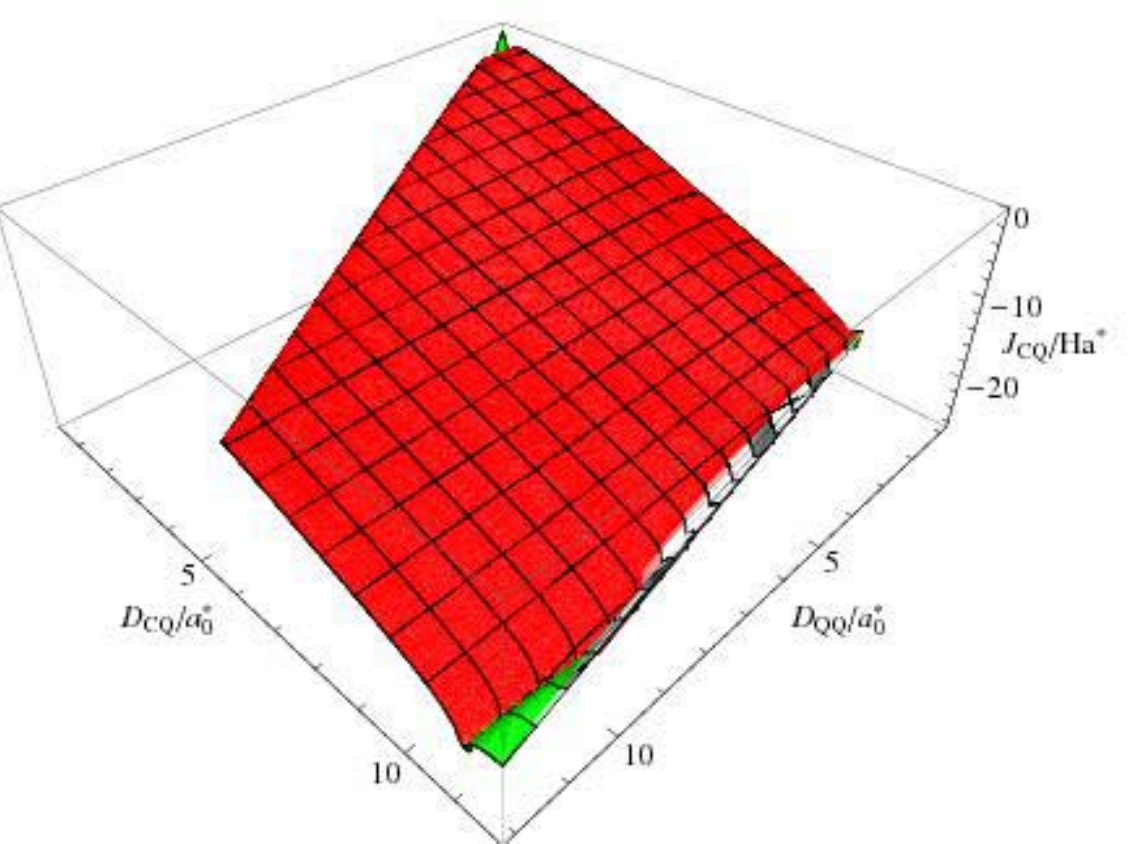}\\
(a)\\
\includegraphics[height=5cm]{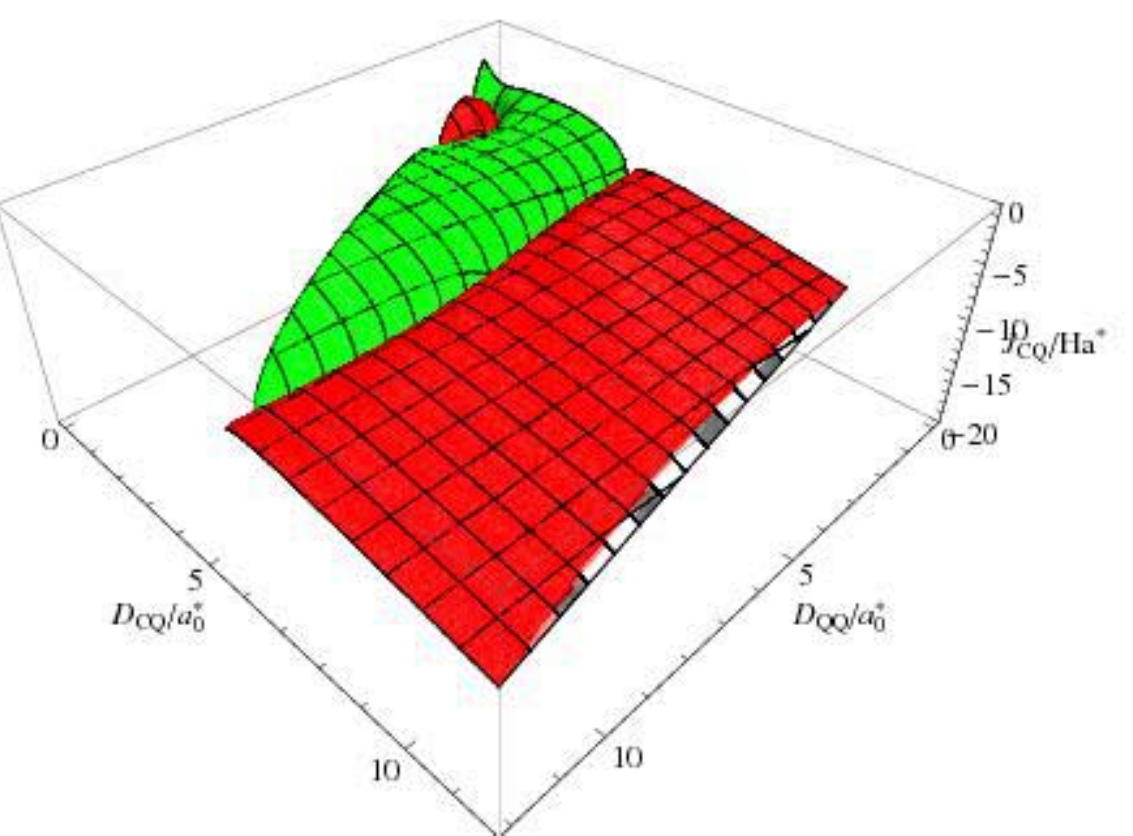}\\
(b)\\
\end{tabular}
\caption{(Colour online.) The control-qubit exchange coupling
$J_{CQ}$ (logarithmic scale) for the case where a shallow-donor
control and deep-donor qubits ($\nu=0.7$) are in the spatial
configuration shown in Figure~\ref{fig:geometry} as a function
of qubit donor distance $D_{QQ}$ and the qubit-control distance
$D_{CQ}$. (a) Control is in the ground state ($\nu=1.0$); (b)
control in the excited state ($\nu=2.0$). The sign of exchange
coupling is coded in color; red: positive (antiferromagnetic) and
green: negative (ferromagnetic). $D_{QQ}$ varies from $1.0$ to $12.0
a_{0}^{*}$; $H$ varies from $0$ to $10 a_{0}^{*}$. Note that the
vertical scales in the two plots are the same.}\label{fig:JCQfig}
\end{figure}

\begin{figure}
\begin{tabular}{c}
\includegraphics[height=5cm]{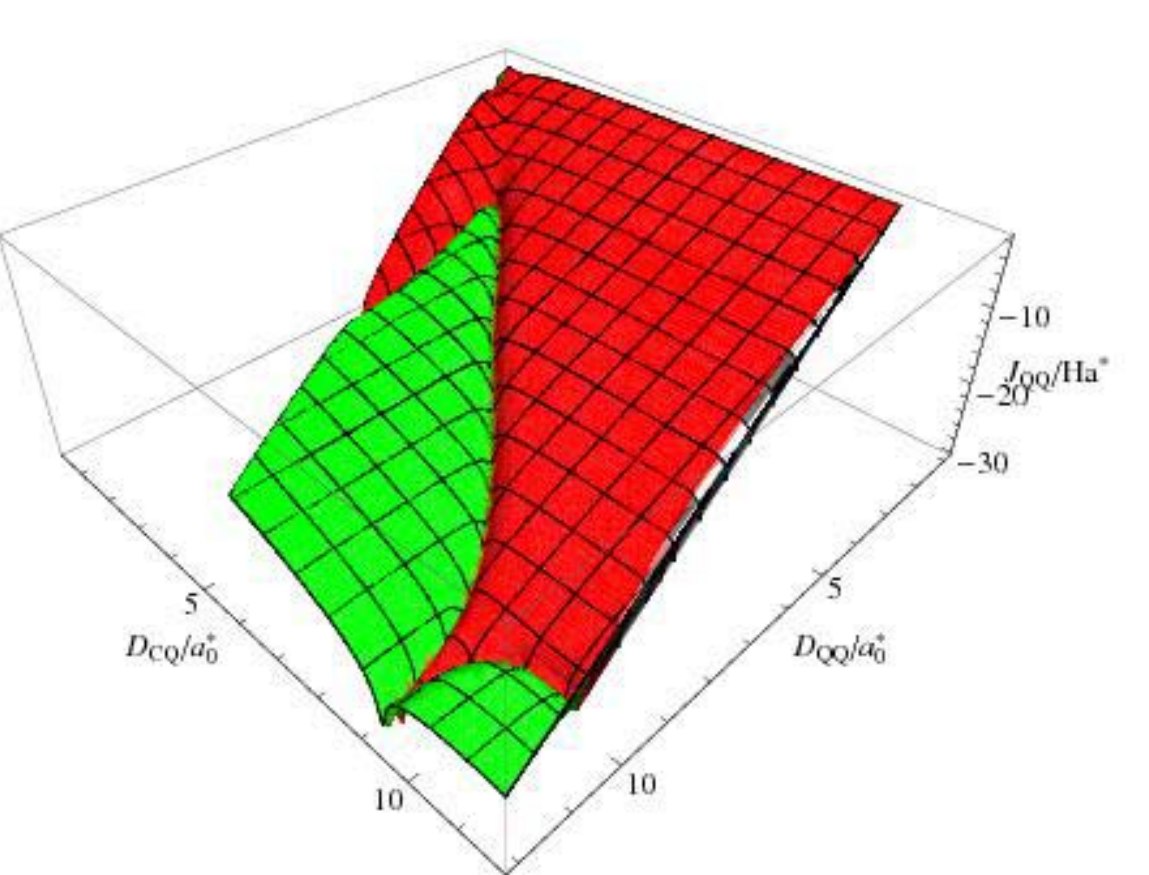}\\
(a)\\
\includegraphics[height=5cm]{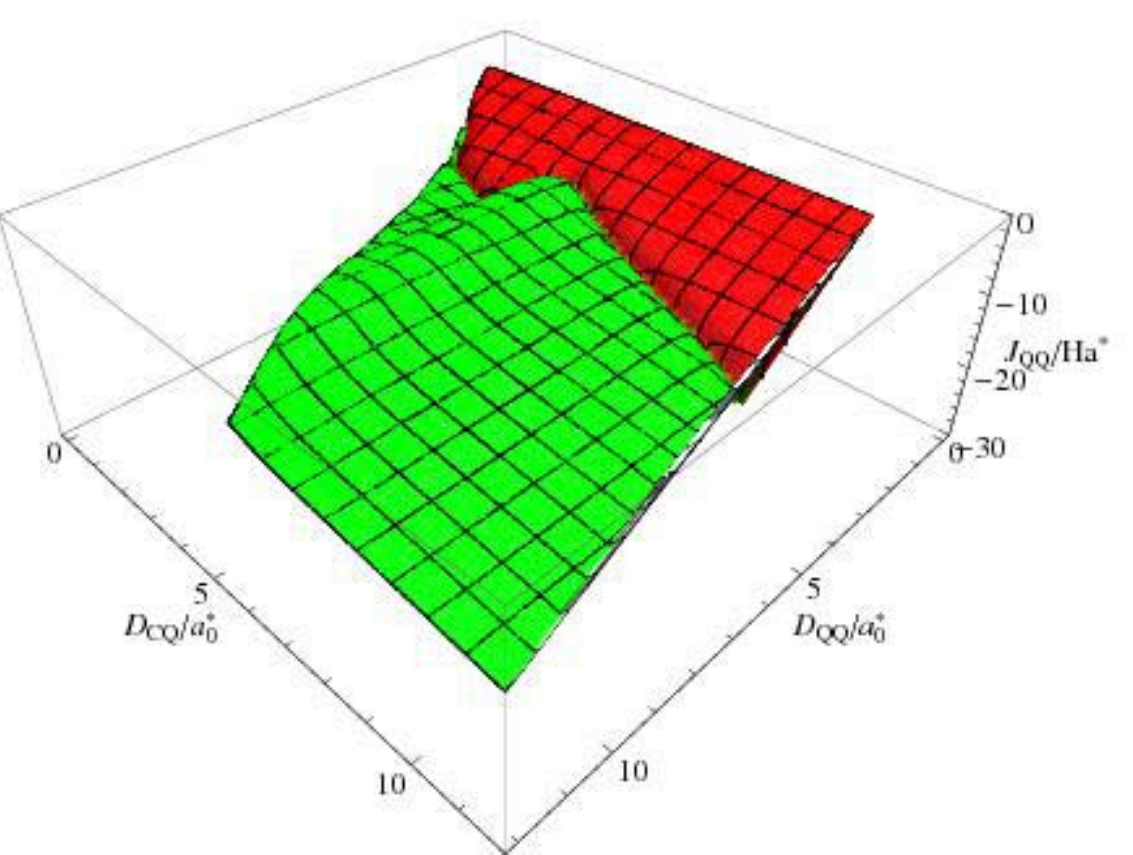}\\
(b)\\
\end{tabular}
\caption{(Colour online.) The qubit-qubit exchange coupling
$J_{QQ}$ (logarithmic scale) for the case where a shallow-donor
control and deep-donor qubits ($\nu=0.7$) are in the spatial
configuration shown in Figure~\ref{fig:geometry} as a function
of qubit donor distance $D_{QQ}$ and qubit-control distance $D_{CQ}$.
(a) control is in the ground state ($\nu=1.0$); (b) control in the
excited state ($\nu=2.0$). The sign of exchange coupling is coded
in color; red: positive (antiferromagnetic) and green: negative
(ferromagnetic). $D_{QQ}$ varies from $1.0$ to $12.0 a_{0}^{*}$; $H$
varies from $0$ to $10 a_{0}^{*}$.  Note that the vertical scales
in the two plots are the same, and that the constraint $D_{CQ}\ge
D_{QQ}$ is enforced by our geometry.}\label{fig:JQQfig}
\end{figure}

In Figures~\ref{fig:JCQfig} and \ref{fig:JQQfig} we show
respectively the control-qubit coupling $J_{CQ}$ and the
qubit-qubit coupling $J_{QQ}$, as the qubit-qubit distance
$D_{QQ}$ varies from $1$ to $12\,a_0^*$ and the height $H$ of the
triangle formed by the qubits and control varies from $0$ to $10
\,a_0^*$. In these and
subsequent figures the geomtry is as shown in
Figure~\ref{fig:geometry}, the two qubits are deep donors, and
the shallow control is either (a) in the ground state ($\nu=1$) or (b) in the
excited state ($\nu=2$).  All results are shown as a function of $D_{QQ}$ and the qubit-control distance $D_{CQ}=\sqrt{H^2+D^2_{QQ}/4}$; the
constraint $D_{CQ}\ge D_{QQ}/2$ is therefore enforced by our
geometry.

It is immediately evident from Figure~\ref{fig:JCQfig}(a) that
even with the control in its ground state, $J_{CQ}$ depends
(weakly) on $D_{QQ}$ as well as (strongly) on $D_{CQ}$. Throughout
most of the region $J_{CQ}$ is positive (antiferromagnetic), as
would be expected for a two-electron system \cite{heisenberg},  although in some
small regions it becomes negative (ferromagnetic).
Figure~\ref{fig:JCQfig}(b), where the control is in the excited
state, shows very different behaviour: now the effective coupling
is much larger and decays much more slowly (as we would expect)
but is also ferromagnetic over a significant part of parameter
space (which is more surprising).  We will see below that this
change is related to a change in the spatial location of the
control electron, as it becomes delocalized over the three
centres.

Similar cooperative behaviour is evident in the behaviour of the
qubit-qubit coupling $J_{QQ}$, which displays a more complex
structure than the exponential decrease with $D_{QQ}$ that one
might naively expect.    Figure~\ref{fig:JQQfig}(a) shows that the
coupling is predominantly exponential in $D_{QQ}$ only when
$D_{QQ}\ll D_{CQ}$, but this behaviour changes when the control atom becomes situated close to the
midpoint of AB.  For $D_{QQ}\ge D_{CQ}$, $J_{QQ}$ is negative
(ferromagnetic) and strongly dependent on $D_{CQ}$ as well as
$D_{QQ}$; this contrasts with the two-electron case \cite{heisenberg}, where
$J_{QQ}$ should be always positive (antiferromagnetic). This
illustrates that even in the ground state the shallow-donor control
electron can cause a significant perturbation to the originally
relatively weak exchange couplings in the two-qubit system.

\begin{figure}
\begin{tabular}{c}
\includegraphics[height=5cm]{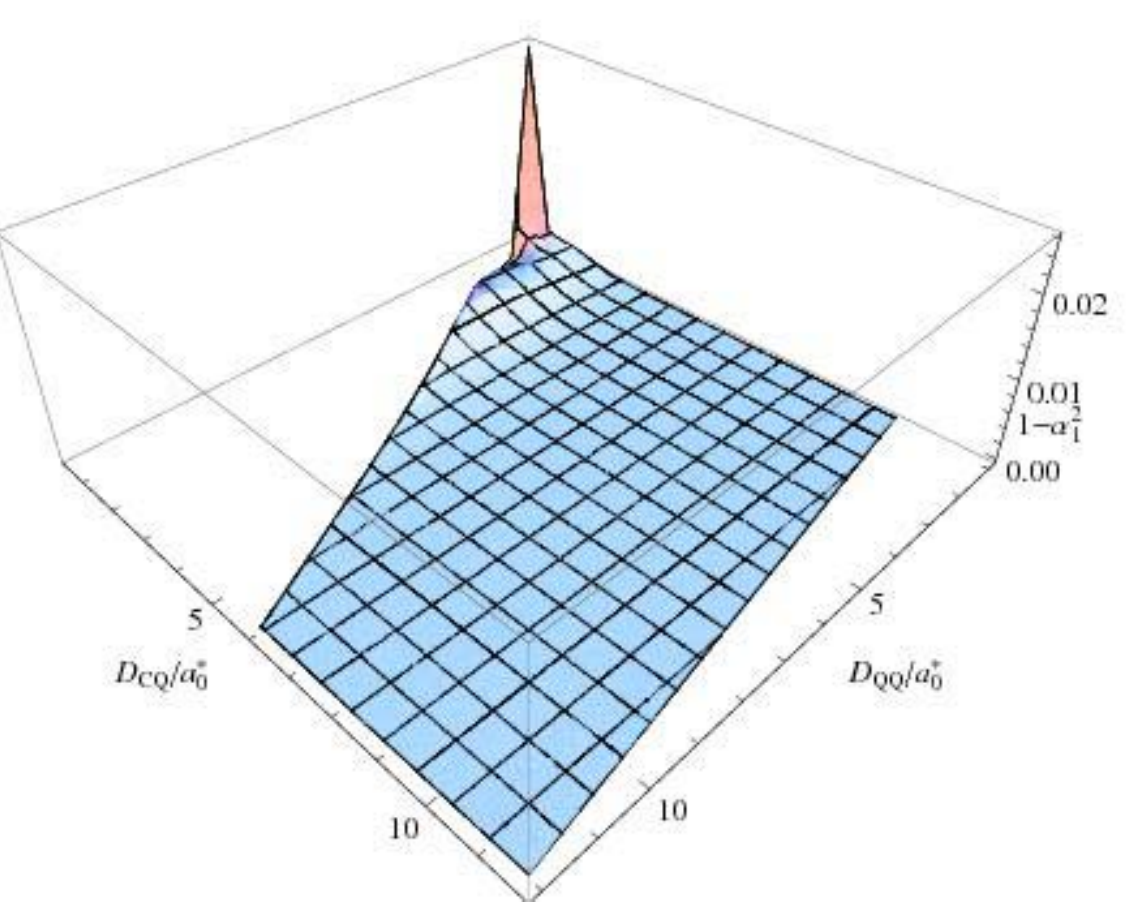}\\
(a)\\
\includegraphics[height=5cm]{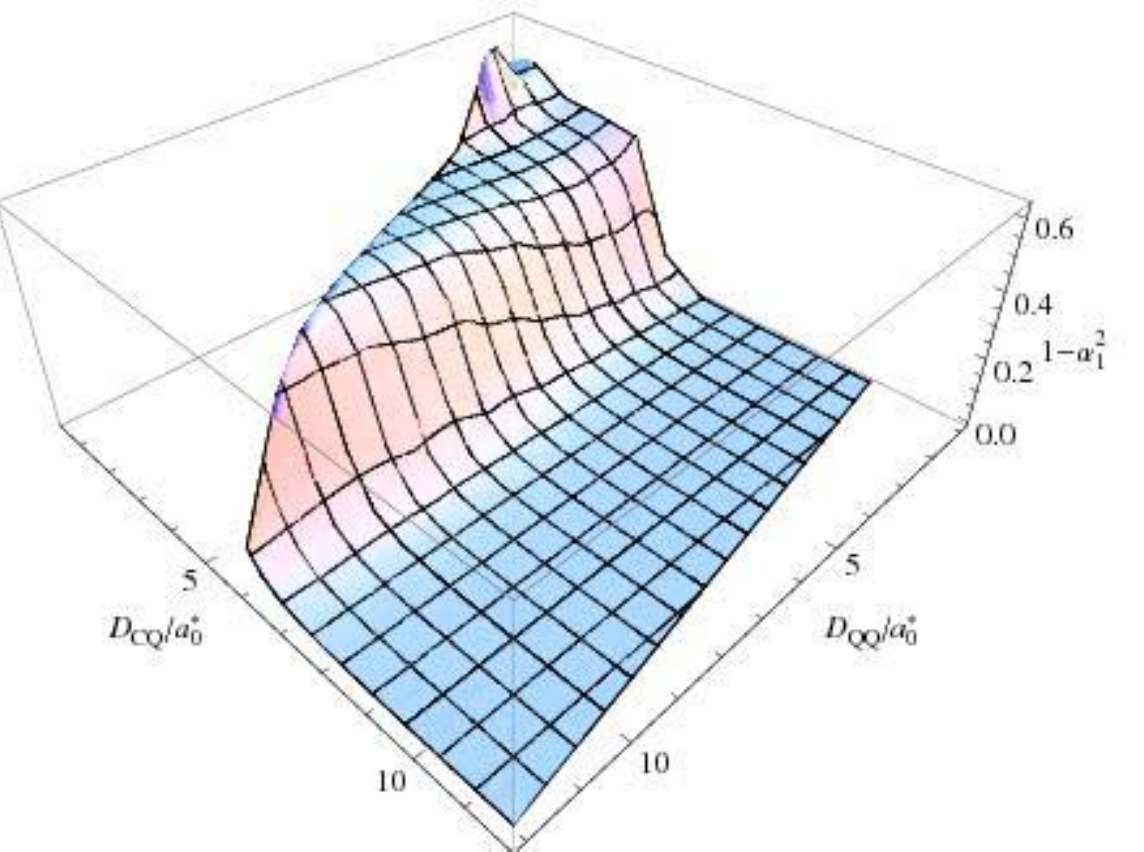}\\
(b)\\
\end{tabular}
\caption{(Colour online.) The probability $1-\alpha_1^2$ of finding
the control donor electron on the qubit cites for the case where
a shallow-donor control and deep-donor qubits ($\nu=0.7$) are in
the spatial configuration shown in Figure~\ref{fig:geometry} as
a function of qubit donor distance $D_{QQ}$ and qubit-control distance $D_{CQ}$. (a) control is in the ground state ($\nu=1.0$); (b)
control in the excited state ($\nu=2.0$). $D_{QQ}$ varies from $1.0$ to
$12.0 a_{0}^{*}$; $H$ varies from $0$ to $10
a_{0}^{*}$.}\label{fig:asqed}
\end{figure}

The transitions between different regimes can be largely
understood from Figure~\ref{fig:asqed}, which shows the
variation of $1-|\alpha|^2$ and hence of the probability (in the
Mulliken sense) that the control electron is located on the qubit sites.  We see that
$|\alpha|^2\approx 1$ throughout, except in the excited state
when the control is relatively close to the qubits.  In that case
$|\alpha|^2\ll 1$ and the control state hybridizes strongly with
the virtual orbitals on the qubit atoms.  In this case the physics
of the exchange is dominated by local interactions between the
control and qubit spins similar to those in the $1s2s$ excited
state of He, where there is a ferromagnetic exchange splitting of
$6,422\,\mathrm{cm}^{-1}$ or $0.80\ \mathrm{eV}$ \cite{woodgate}.

\begin{figure}
\begin{tabular}{c}
\includegraphics[height=5cm]{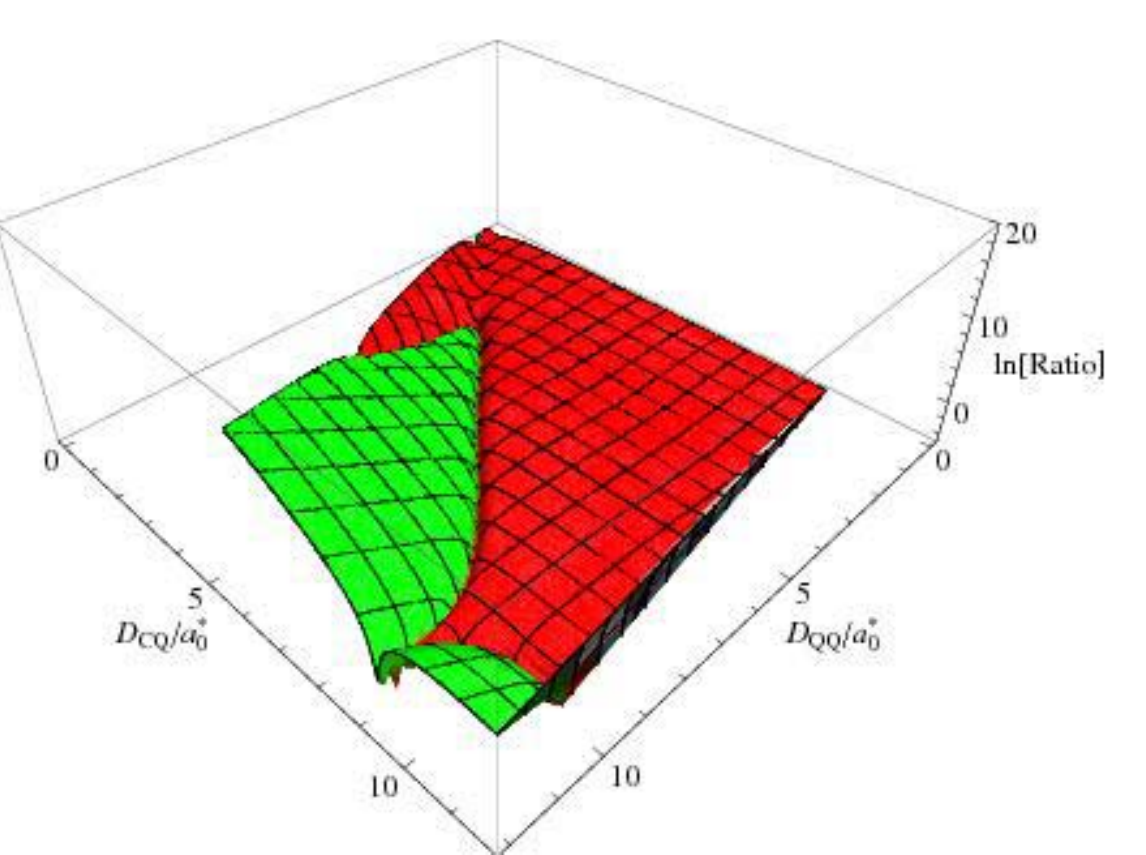}\\
(a)\\
\includegraphics[height=5cm]{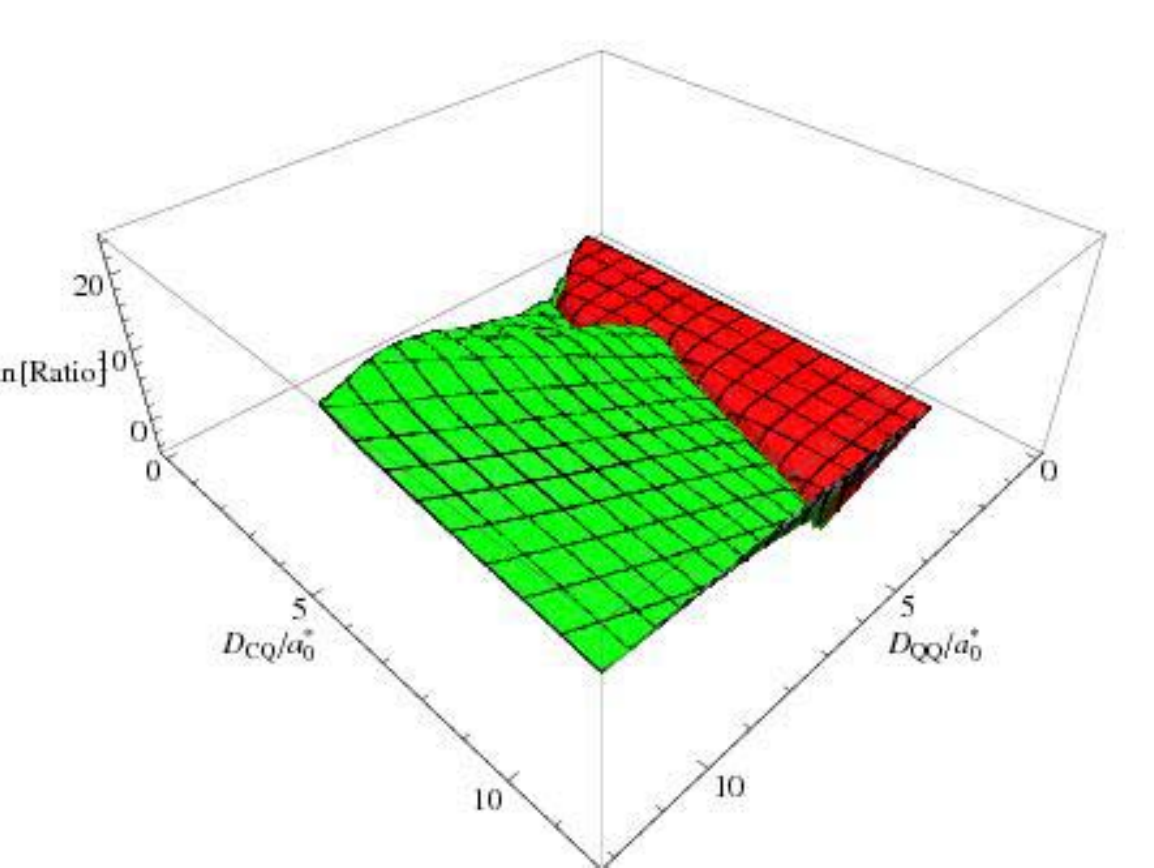}\\
(b)\\
\end{tabular}
\caption{(Colour online.) The ratio (logarithmic scale) of
the qubit-qubit exchange couplings $J_{QQ}$ for two deep-donor qubits ($\nu=0.7$) with and without a shallow-donor control.
The spatial configuration is as shown in Figure~\ref{fig:geometry}; results are shown as a function of the qubit-qubit distance $D_{QQ}$ and the
qubit-control distance $D_{CQ}$.  In (a) the control is in the ground state
($\nu=1.0$); in (b) the control in the excited state ($\nu=2.0$). The
sign of the ratio is coded in color; red: positive (antiferromagnetic interaction in presence of control), and green: negative (ferromagnetic). In the absence of the control, the interaction is always antiferromagnetic.
\label{fig:ratiojqqjqq0}}
\end{figure}

We focus on how the control electron alters $J_{QQ}$ and plot the ratio of this quantity with and without the control present in
Figure~\ref{fig:ratiojqqjqq0}. As expected, when the control is very
far away from qubits ($D_{CQ}\gg D_{QQ}$), the ratio is close to unity.  When this condition is not satisfied (i.e., for $D_{CQ}\ltsimeq D_{QQ}$), $J_{QQ}$ is strongly modified and (over a large part of the region) ferromagnetic.  In the excited state (Figure~\ref{fig:ratiojqqjqq0}(b)) the coupling is enhanced and ferromagnetic almost everywhere, unless $D_{CQ}\gg D_{QQ}$.  We will show in \S\ref{sec:ringexchange} that the ferromagnetic interactions are a consequence of ring exchange among the three atoms.

\begin{figure}
\begin{tabular}{c}
\includegraphics[height=5cm]{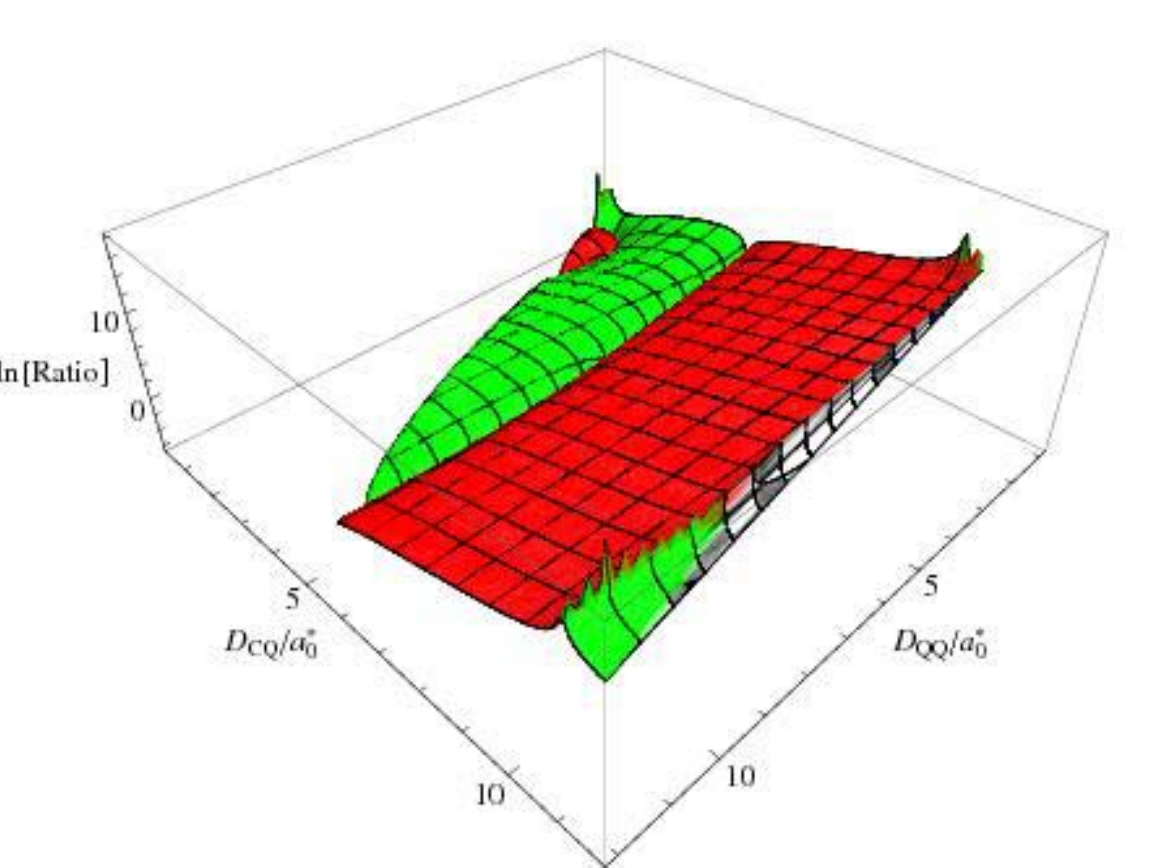}\\
(a)\\
\includegraphics[height=5cm]{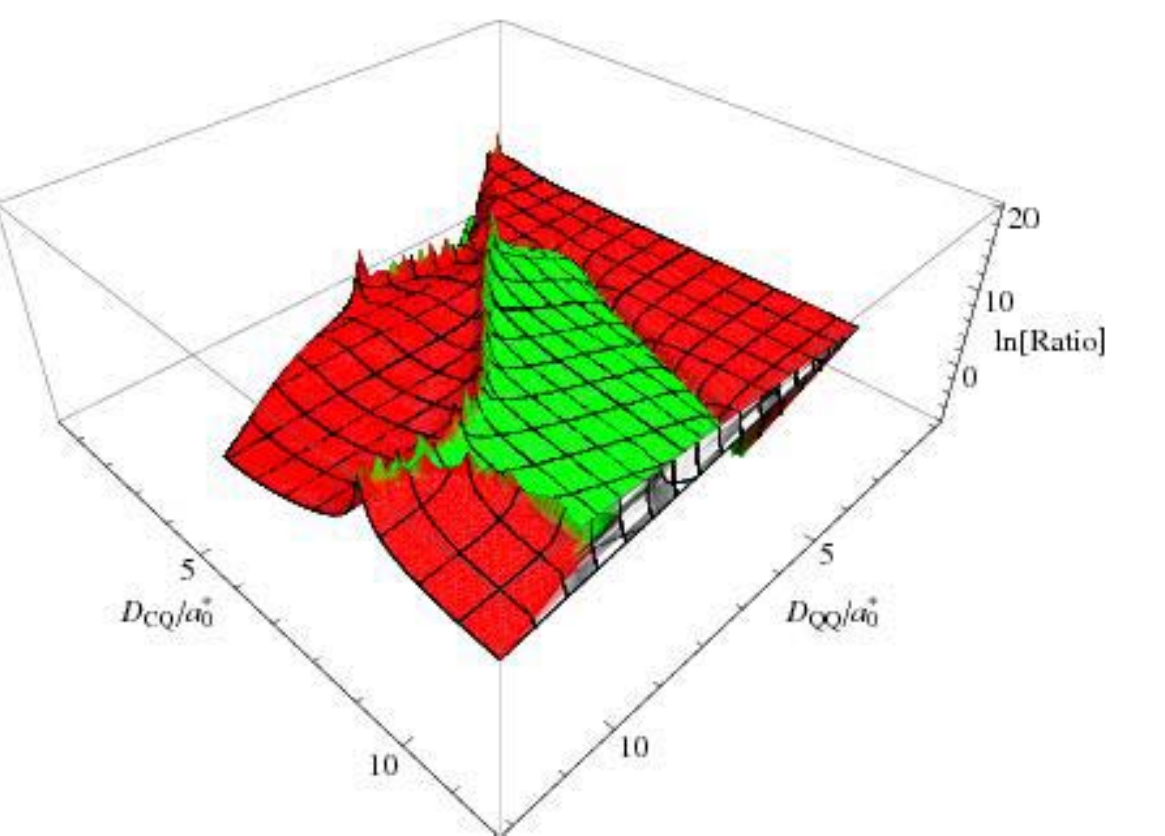}\\
(b)\\
\end{tabular}
\caption{(Colour online.) The ratios ${J^e}/{J^g}$
(logarithmic scale) of the exchange interactions for the control excited ($J^e$, with $\nu=2.0$) to control-unexcited ($J^g$, with $\nu=1.0$)
 for a shallow-donor control and
deep-donor qubits: (a) control-qubit exchange $J_{CQ}$, (b) qubit-qubit exchange $J_{QQ}$.
The sign of the ratio is coded in color: red: positive and
green: negative.}\label{fig:ratioonoff}
\end{figure}

Finally, in Figure~\ref{fig:ratioonoff}, we focus on the changes brought about by excitation (the `on/off ratios' of the exchange).
As would be expected, excitation almost always increases the magnitude of the exchange; Figure~\ref{fig:ratioonoff}(a) (for $J_{QQ}$) shows that this effect is  largest for large $D_{QQ}$ and moderate $D_{CQ}$, although there is a relatively complicated behaviour of the sign of the ratio as a result of the differing crossovers between ferromagnetic and antiferromagnetic exchange in the ground and excited states (Figure~\ref{fig:JQQfig}).  Figure~\ref{fig:ratioonoff}(b) (for $J_{CQ}$) shows a simpler pattern: the on/off ratio increases with $D_{CQ}$ and remains only weakly dependent on $D_{QQ}$.

\section{Ring-exchange model}\label{sec:ringexchange}

\subsection{Multi-center ring exchange}
In this section, we present an alternative calculation of the exchange
constants by using Green's-function perturbation theory. This approach has the advantage that it gives a clear picture of the origins of different contributions to the exchange.  In particular it allows us to
understand the origin of the ferromagnetism of $J_{QQ}$ in
\S\ref{sec:results}: owing to the presence of the shallow control
atom, multi-body ring exchange process \cite{roger83} become possible and higher-order ring exchange process (which can be ferromagnetic) may dominate over the direct second-order exchange between the qubits provided $D_{CQ}\ll D_{QQ}$.

\begin{figure}[htbp]
\begin{center}
\begin{tabular}{ccc}
\includegraphics[width=2.5cm,height=1.5cm]{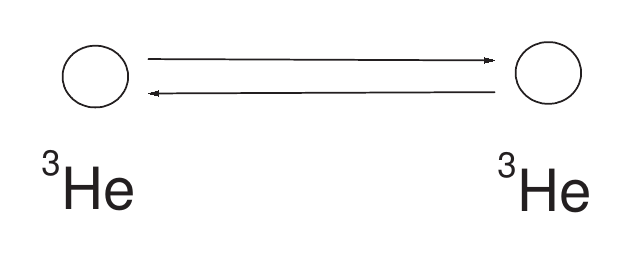}&
\includegraphics[width=2.5cm,height=3cm]{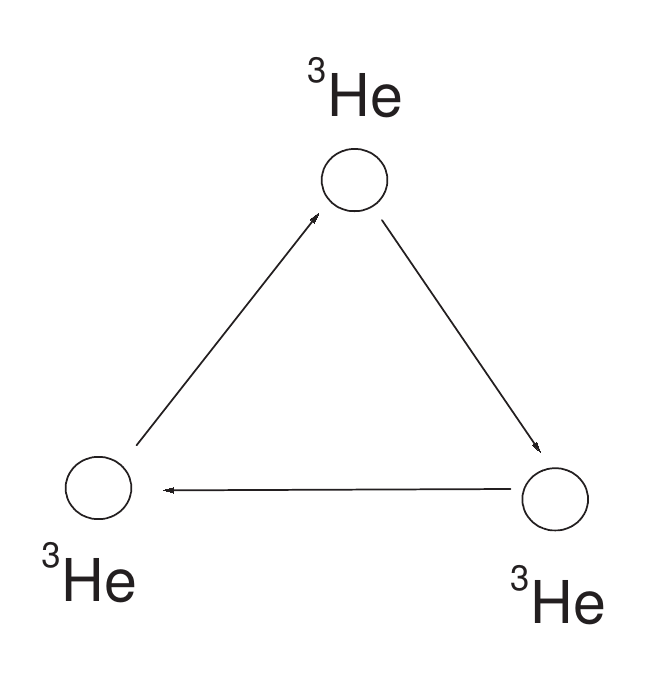}&
\includegraphics[width=2.5cm,height=3cm]{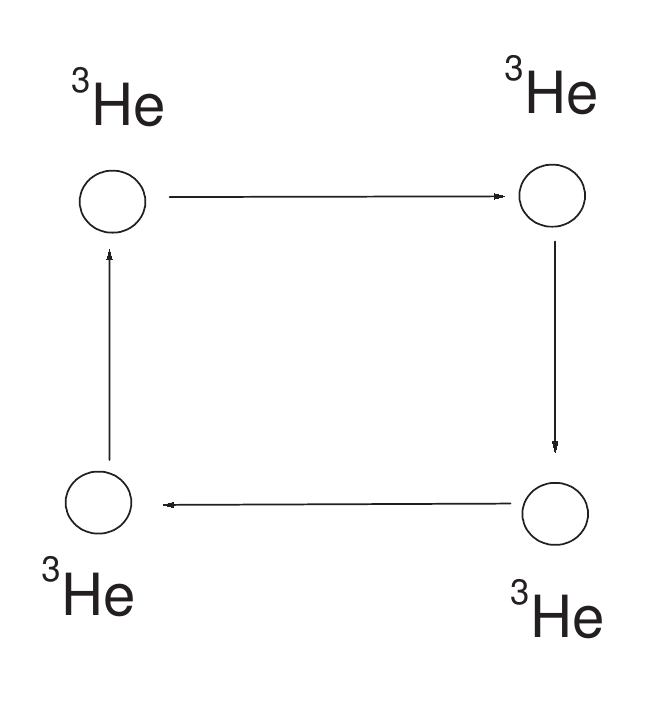}\\
(a)&(b)&(c)
\end{tabular}
\end{center}
\caption{Schematic view of possible exchange couplings between
helium atoms in solid $^3\mathrm{He}$: (a) conventional two-body
exchange (antiferromagnetic); (b) three-body ring exchange
(ferromagnetic); (c) Four-body ring exchange (anti-ferromagnetic).}\label{pic:234th}
\end{figure}

The essential physics of our system is similar to that in solid
$^3\mathrm{He}$ \cite{roger83}, despite the very different chemical nature of the material.  In solid
$^3\mathrm{He}$, two-body exchange always leads to a Heisenberg
Hamiltonian with antiferromagnetic exchange (Figure
\ref{pic:234th}a), while three-body ring exchange (Figure
\ref{pic:234th}b) and four-body ring exchange (Figure
\ref{pic:234th}c) lead respectively to ferromagnetic and anti-ferromagnetic contributions.

\begin{figure}[htbp]
\begin{center}
\begin{tabular}{cc}
\includegraphics[width=4cm,height=4cm]{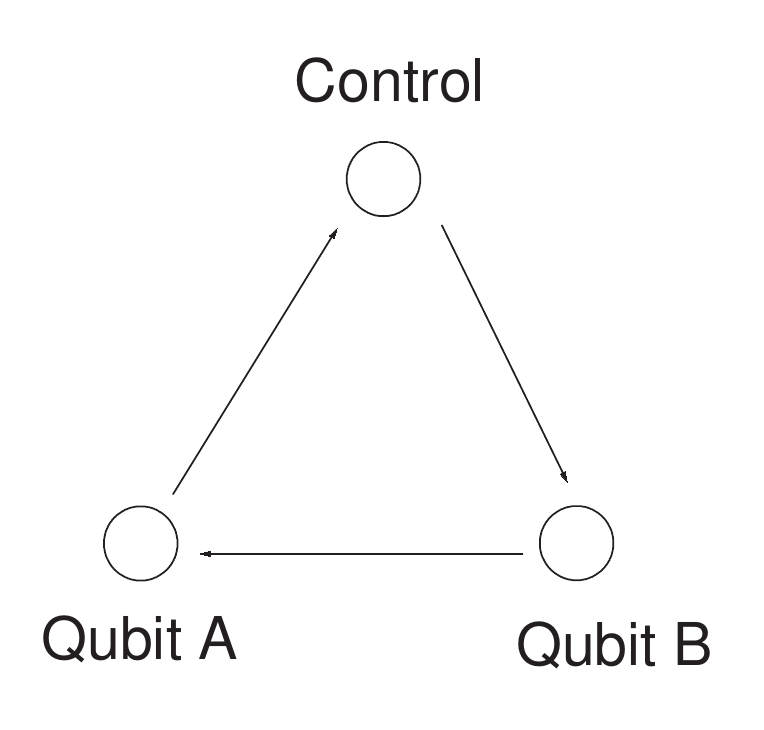}&
\includegraphics[width=4cm,height=4cm]{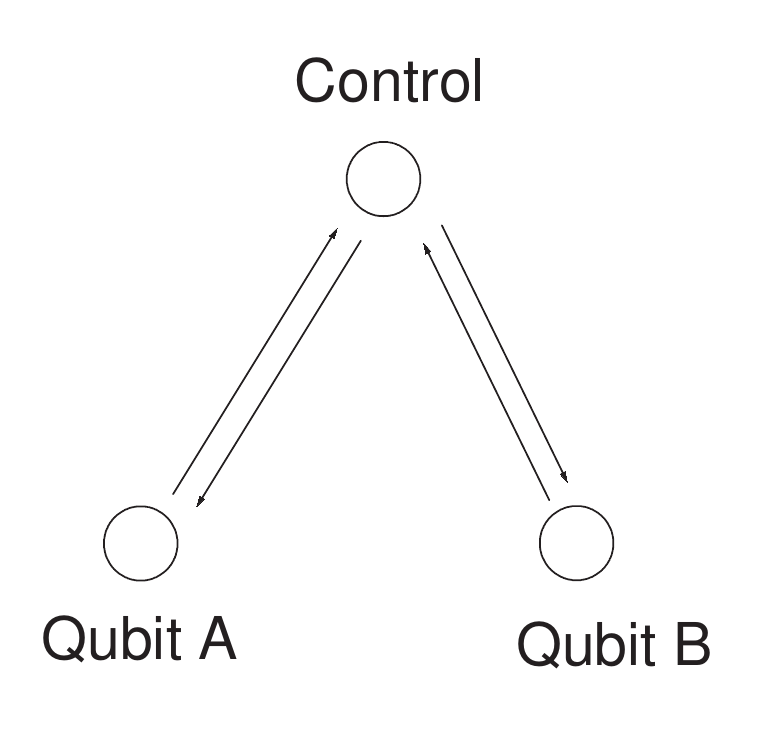}\\
(a)&(b)
\end{tabular}
\end{center}
\caption{Schematic view of multi-centre exchange processes in the
control-qubit system: (a) third-order exchange scheme; (b)
fourth-order exchange scheme.}\label{pic:3rd4th}
\end{figure}

As in the solid $^3\mathrm{He}$ system we need to include
third-order and fourth-order processes (Figure \ref{pic:3rd4th})
in our Green's function perturbation theory calculation in order
to capture these effects of control atom.

\subsection{Green's functions time-independent perturbation
theory}\label{sec:greenfunction}
Suppose the Hamiltonian for a
system can be separated into an unperturbed part of $H_0$, and a
perturbation $V$; $H=H_0+V$. We assume the eigenvalues and
eigenfunctions of $H_0$ are easily obtained. Green's functions
$G_{0}(z)$ and $G(z)$ corresponding to $H_0$ and $H$,
respectively, are $G_0(z)=(z-H_0)^{-1}$, and $G(z)=(z-H)^{-1}$.
$G$ and $G_0$ are connected by the Dyson equation
\begin{eqnarray}\label{eq:dyson}
G(z)&=& [1-G_0(z)V]^{-1}G_0(z)
\\&=& G_0+G_0VG_0+G_0VG_0VG_0+\ldots,
\end{eqnarray}
and the effective Hamiltonian which can mix a particular subset $\mathcal{S}$ of the eigenstates of $\hat{H}_0$ is
\begin{eqnarray}\label{eq:energyshift}
\Delta \hat{H}_{\mathrm{eff}}(z)&=&
\hat{P}\hat{V}\hat{P}+\hat{P}V\hat{Q}\hat{G}_0(z)\hat{Q}V\hat{P}\nonumber
\\&&+\hat{P}V\hat{Q}\hat{G}_0(z)\hat{Q}V\hat{Q}\hat{G}_0(z)\hat{Q}V\hat{P}+\ldots,
\end{eqnarray}
where $\hat{Q}=\hat{1}-\hat{P}$, $\hat{P}$ is the projection operator onto the set $\mathcal{S}$,
and the 1st-, 2nd-, 3rd-order... energy shifts are adumbrated on the right-hand side of equation~(\ref{eq:energyshift}).
In this formula, $\Delta \hat{H}_{\mathrm{eff}}$ depends on $z$,
which should therefore be chosen to correspond to the energy of the states
$\mathcal{S}$. In the next section, we will show how we use
Green's function perturbation theory to extract the exchange
constants.

\subsection{Simple model of the three-centre problem}\label{sec:
simplemodelfor3centre}
From Figure~\ref{fig:asqed} we see that the control electron resides
on the control atom over the majority of parameter space; nevertheless, Figure~\ref{fig:JQQfig} shows $J_{QQ}$ is
ferromagnetic for much of this region. Therefore for simplicity we ignore the
donor excited states, retaining only one orbital per site. We consider as parts of $\hat{H}_0$ all processes which do not transfer charge between centres (both
single-electron terms and Coulomb interactions within one atom, and Coulomb interactions between different centres);
the perturbation $V$ then contains hopping operators that transfer electrons between centres.  Equation~(\ref{eq:dyson}) then has a simple interpretation reflected in (\ref{eq:energyshift}): $G_0$ gives the propagation of uncoupled atoms, while $\hat{V}$ exchanges electrons between those atoms.

\begin{figure}[htbp]
\begin{center}
\begin{tabular}{ccc}
\includegraphics[height=2cm]{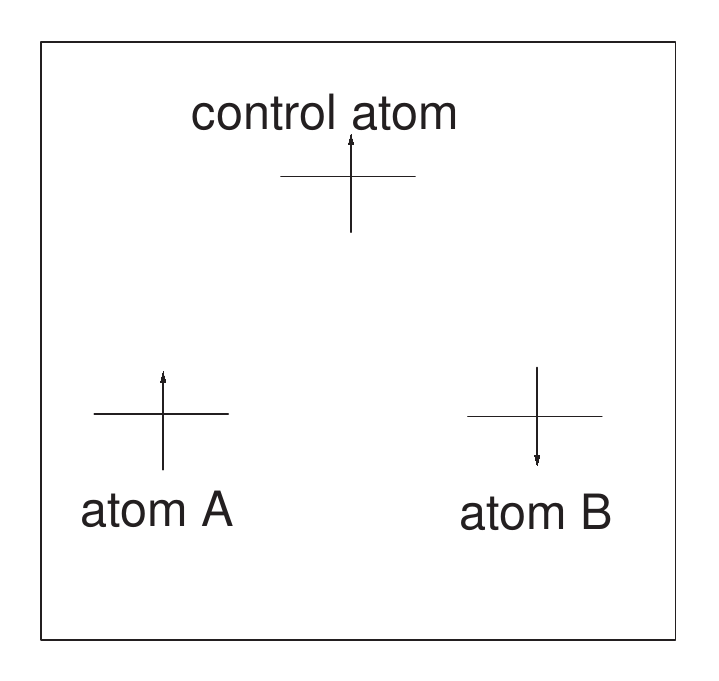}&
\includegraphics[height=2cm]{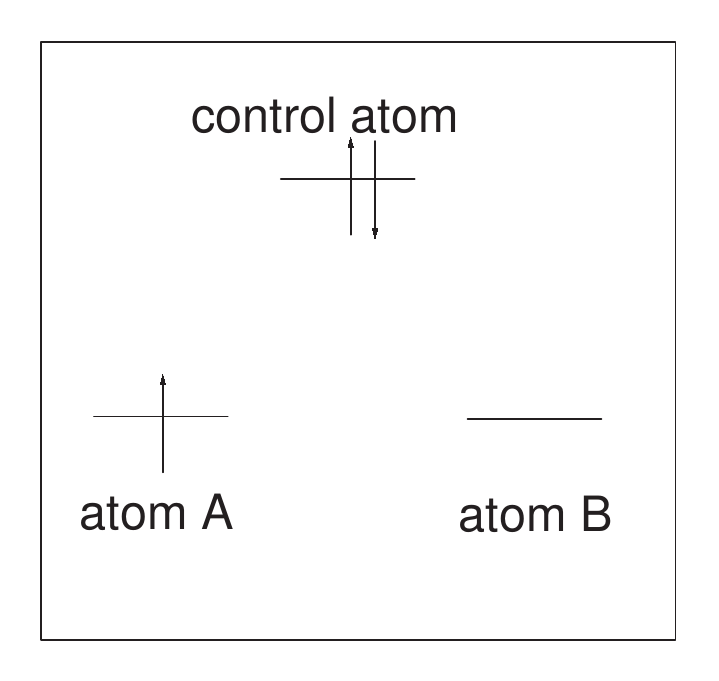}&
\includegraphics[height=2cm]{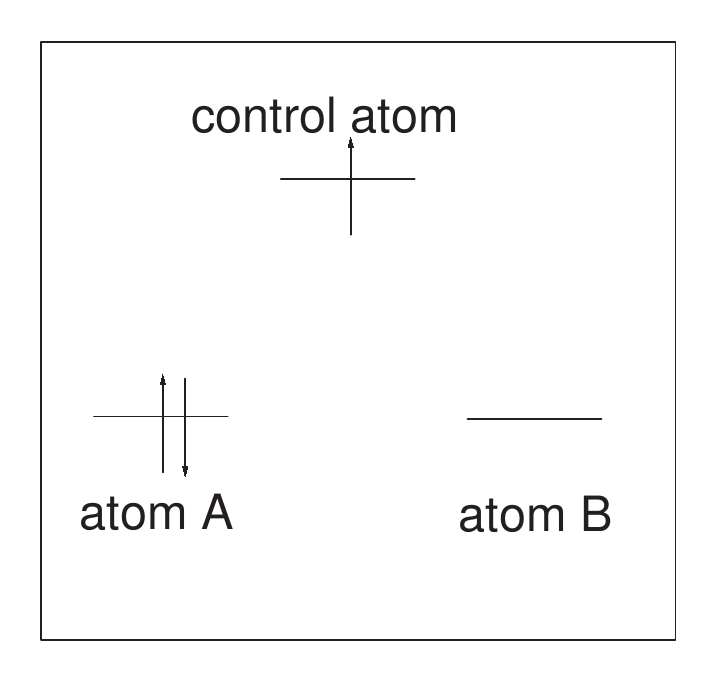}\\
$\ket{1}=c^\dagger_{A\uparrow}c^\dagger_{C\uparrow}c^\dagger_{B\downarrow}\ket{0}$
&$\ket{2}=c^\dagger_{A\uparrow}c^\dagger_{C\uparrow}c^\dagger_{C\downarrow}\ket{0}$
&$\ket{3}=c^\dagger_{A\uparrow}c^\dagger_{A\downarrow}c^\dagger_{C\uparrow}\ket{0}$\\
\includegraphics[height=2cm]{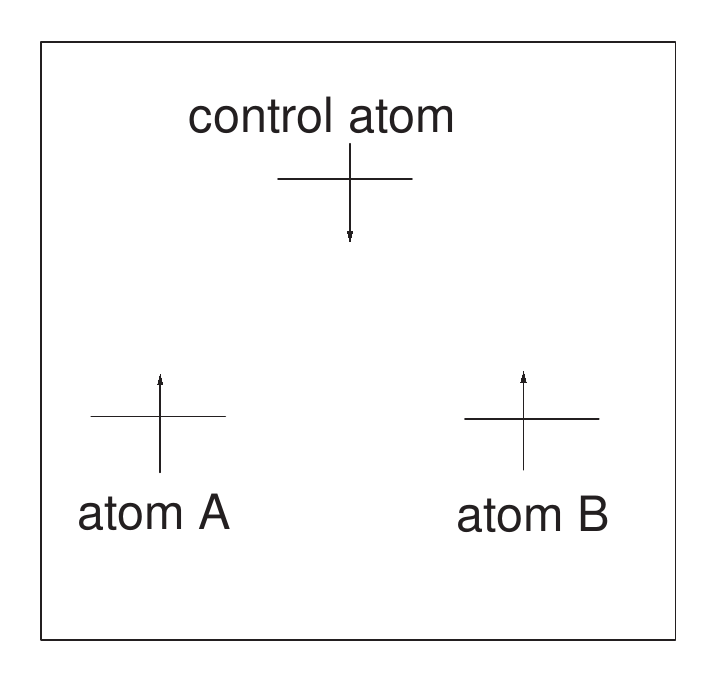}&
\includegraphics[height=2cm]{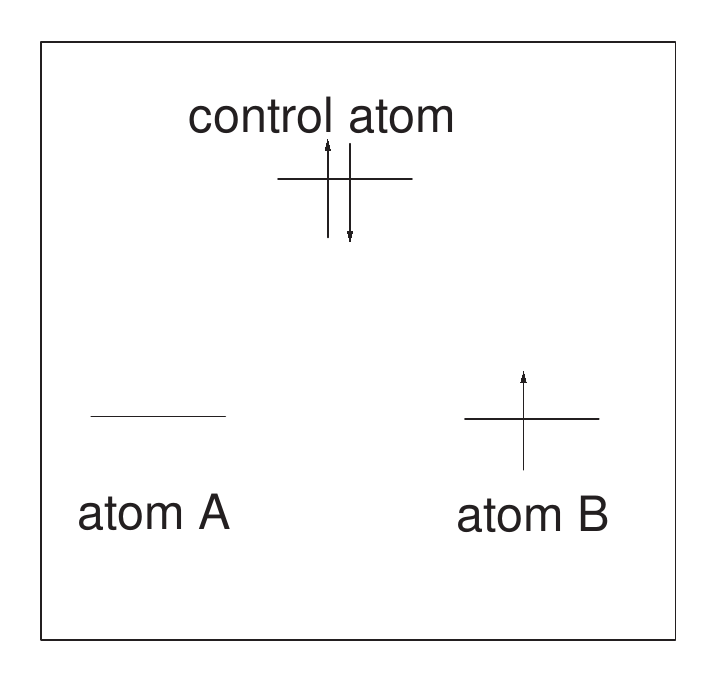}&
\includegraphics[height=2cm]{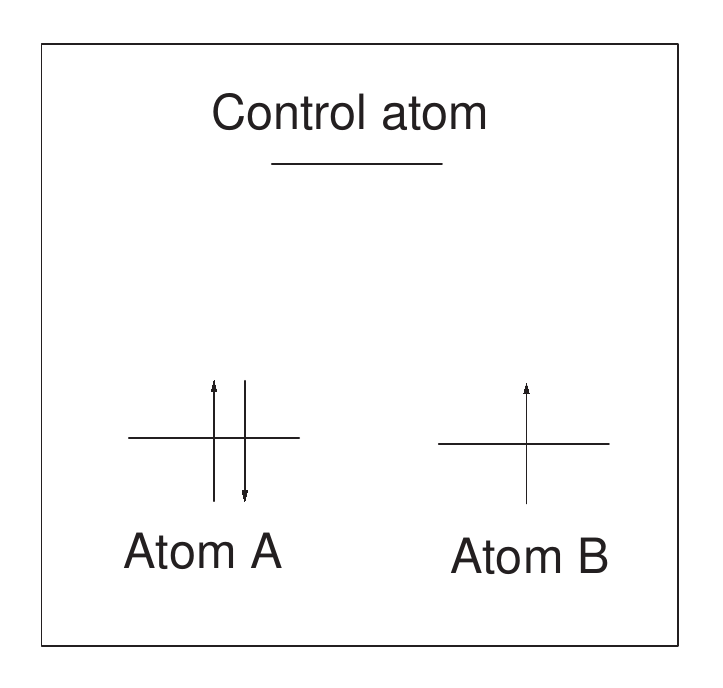}\\
$\ket{4}=c^\dagger_{A\uparrow}c^\dagger_{C\downarrow}c^\dagger_{B\uparrow}\ket{0}$
&$\ket{5}=c^\dagger_{C\uparrow}c^\dagger_{C\downarrow}c^\dagger_{B\uparrow}\ket{0}$
&$\ket{6}=c^\dagger_{A\uparrow}c^\dagger_{A\downarrow}c^\dagger_{B\uparrow}\ket{0}$\\
\includegraphics[height=2cm]{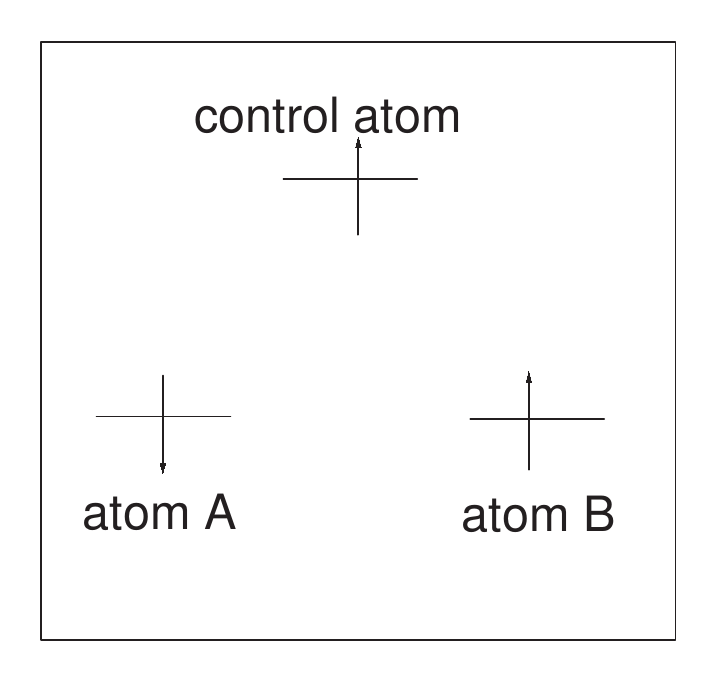}&
\includegraphics[height=2cm]{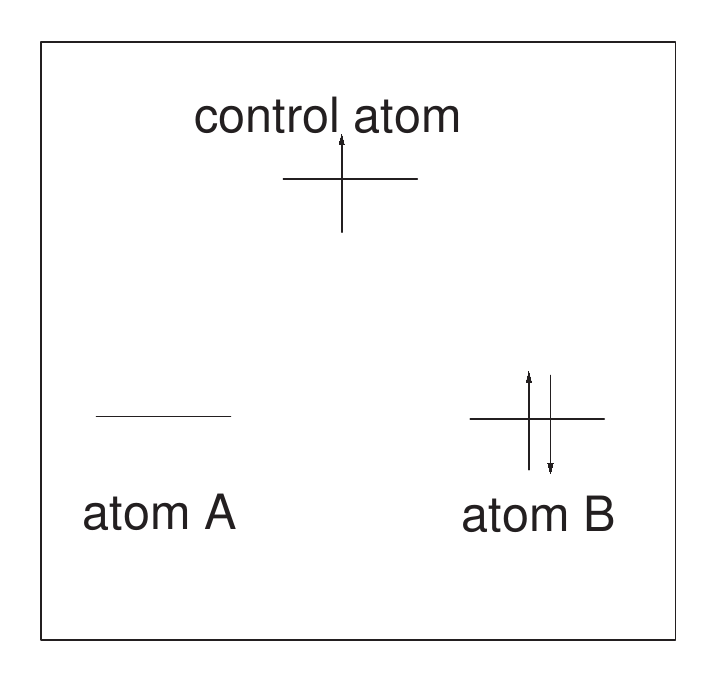}&
\includegraphics[height=2cm]{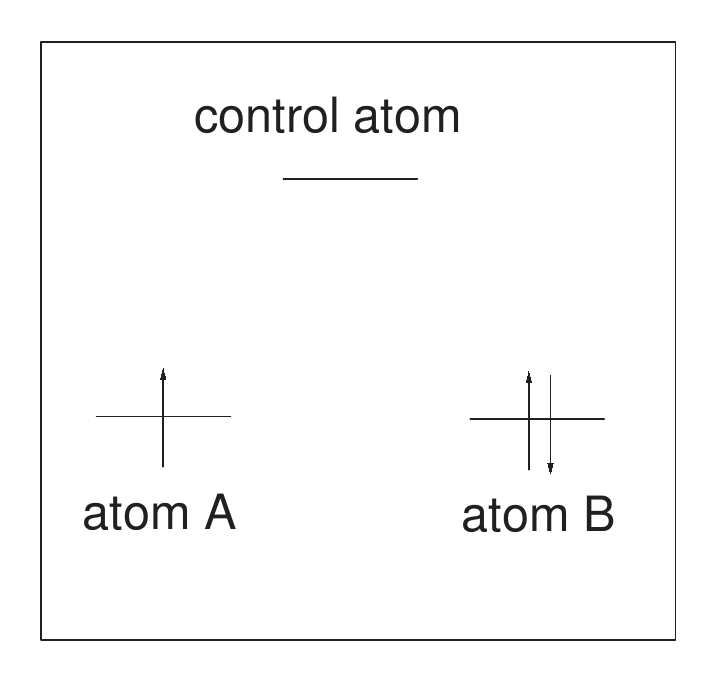}\\
$\ket{7}=c^\dagger_{A\downarrow}c^\dagger_{C\uparrow}c^\dagger_{B\uparrow}\ket{0}$
&$\ket{8}=c^\dagger_{C\uparrow}c^\dagger_{B\uparrow}c^\dagger_{B\downarrow}\ket{0}$
&$\ket{9}=c^\dagger_{A\uparrow}c^\dagger_{B\uparrow}c^\dagger_{B\downarrow}\ket{0}$\\
\end{tabular}
\end{center}
\caption{Schematic of nine possible states according to the calculation procedure in \S\ref{sec:
simplemodelfor3centre}. We define second-quantifying operators:
$c^\dagger_{A\sigma}$, $c_{A\sigma}$, $c^\dagger_{C\sigma}$,
$c_{C\sigma}$, $c^\dagger_{B\sigma}$, $c_{B\sigma}$, where
$\sigma$ is $\uparrow$ or $\downarrow$, and $c^\dagger_{A\sigma}$
is electron-creating operator on the state of qubit $A$,
etc.}\label{pic:9states}
\end{figure}

Then we carry out our calculation in the following steps.
First, we select a particular eigenvalue of the $z$-component of the total spin $S_z$ (in the following discussion,
we choose $S_z=\frac{1}{2}$) ($S_z$ is a good
quantum number since we ignore the spin-orbit coupling).
Next we enumerate all the possible configurations for this $S_z$.
(For $S_z=\frac{1}{2}$ the nine configurations are shown in
Figure~(\ref{pic:9states}).)  We choose which of these states we will retain as our set $\mathcal{S}$; typically, this is the low-energy subspace with one electron per donor, so that Coulomb intreactions are minimized.
Next we construct the matrices of $G_0$ and $V$ based on all these possible states (see \S\ref{sec:greenfunction}), and use these matrices to derive  2nd-, 3rd-, and 4th-order terms in $\Delta{H}_{\mathrm{eff}}$ according to
equation~(\ref{eq:energyshift}).  (We suppose the higher terms are negligible.)
Finally, as in the variatonal calculation (\S\ref{sec:variational}), we identify the values $J_{CQ}$ and $J_{QQ}$ by comparing the effective Hamiltonian in the last step
with the Heisenberg Hamiltonian (\ref{eq:exchangeH}).

\subsection{Perturbation theory calculation}
\subsubsection{The unperturbed hamiltonian $H_{0}$.}
We take for $H_0$ a simple extended Hubbard-type Hamiltonian:
\begin{eqnarray}
\hat{H}_0&=&E_C\hat{n}_C+E_Q(\hat{n}_A+\hat{n}_B)\nonumber\\
&&+U_C\hat{n}_{C\uparrow}\hat{n}_{C\downarrow}+U_Q(\hat{n}_{A\uparrow}\hat{n}_{A\downarrow}+\hat{n}_{B\uparrow}\hat{n}_{B\downarrow})\nonumber\\
&&+V_{QQ}\hat{n}_A\hat{n}_B+V_{CQ}\hat{n}_C(\hat{n}_A+\hat{n}_B).
\end{eqnarray}
Here $E_{C}$ and $E_Q$ are the single-particle energies of the control
and qubit atoms, $U_C$ and
$U_Q$ are on-site Coulomb interactions in the control and qubit, and $V_{QQ}$ and $V_{CQ}$ are respectively Coulomb
interactions between electrons in the two qubit sites, and between
electrons in the control and qubit sites.  $\hat{n}_{i\sigma}$ is the number operator for electrons with spin $\sigma$ on site $i$, and $\hat{n}_i=\hat{n}_{i\uparrow}+\hat{n}_{i\downarrow}$.

We take for the perturbation $\hat{V}$ the hopping terms that transfer electrons
respectively between the control atom and the
qubits, and from qubit to qubit.  If we assume the corresponding amplitudes $t_{CQ}, t_{QQ}$ are real, then we can write
\begin{eqnarray}
\hat{V}&=&\sum_\sigma[t_{CQ}(\hat{c}^\dagger_{A\sigma}\hat{c}_{C\sigma}+\hat{c}^\dagger_{C\sigma}\hat{c}_{A\sigma}+\hat{c}^\dagger_{B\sigma}\hat{c}_{C\sigma}
+\hat{c}^\dagger_{C\sigma}\hat{c}_{B\sigma})
\nonumber\\&&+t_{QQ}(\hat{c}^\dagger_{A\sigma}\hat{c}_{B\sigma}+\hat{c}^\dagger_{B\sigma}\hat{c}_{A\sigma})],
\end{eqnarray}
where $\hat{c}_{i\sigma}$ annihilates an electron with spin $\sigma$ at site $i$.

\subsubsection{Perturbation theory and reproduction of the Heisenberg
spin Hamiltonian}

Now we have $G_0$ and $V$, we can reproduce the Heisenberg
Hamiltonian from the 2nd-, 3rd- and 4th-order perturbations to
find the corresponding contributions to $J_{QQ}$ and $J_{CQ}$
according the calculation procedure in \S\ref{sec:
simplemodelfor3centre}:
\begin{eqnarray}
J_{QQ}\simeq J_{QQ}^{(2)}+J_{QQ}^{(3)}+J_{QQ}^{(4)}\\
J_{CQ}\simeq J_{CQ}^{(2)}+J_{CQ}^{(3)}+J_{CQ}^{(4)}.
\end{eqnarray}
$J_{QQ}^{(i)}$, $J_{CQ}^{(i)}$, $(i=2,3,4)$ are functions of
$V_{QQ}$, $V_{CQ}$, $U_Q$, $U_C$, $t_{QQ}$, and $t_{CQ}$; all the conributions are given explicitly in Appendix~\ref{sec:fullptresults}.

Motivated by the known physics of the defect problem, we choose suitable
values for these variables to illustrate our results.  We expect the hopping integrals $t_{QQ}$ and  $t_{CQ}$ to decay exponentially with $D_{QQ}$ and $D_{CQ}$ respectively, at rates determined by the sum of the orbital exponents of the relevant atom pairs.  We also expect $V_{QQ},V_{CQ}<U_{C}<U_{Q}$ because the on-site Coulomb
interaction should be larger than the Coulomb interaction between
different centres, and the
qubit spatial state is more localized than the control state.

\begin{figure}[htbp]
\begin{tabular}{c}
\includegraphics[height=5cm]{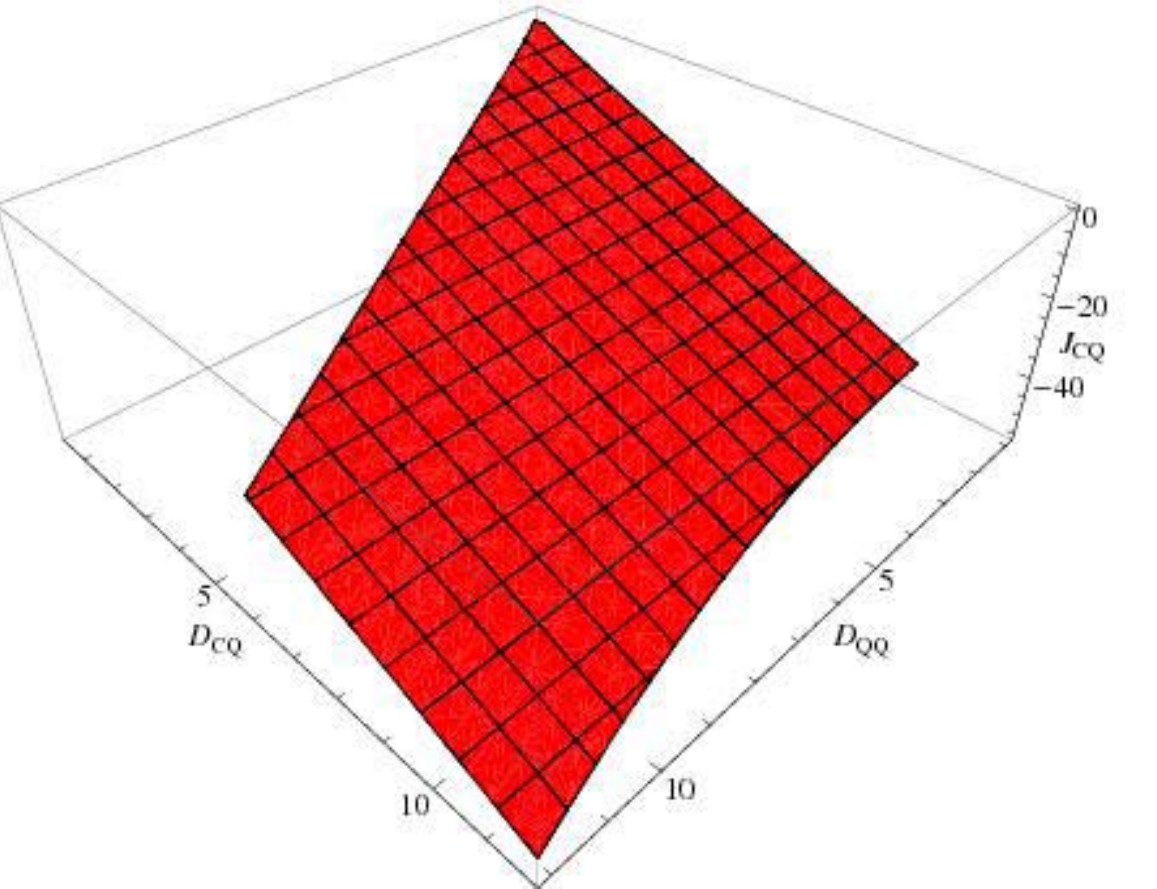}\\
(a)\\
\includegraphics[height=5cm]{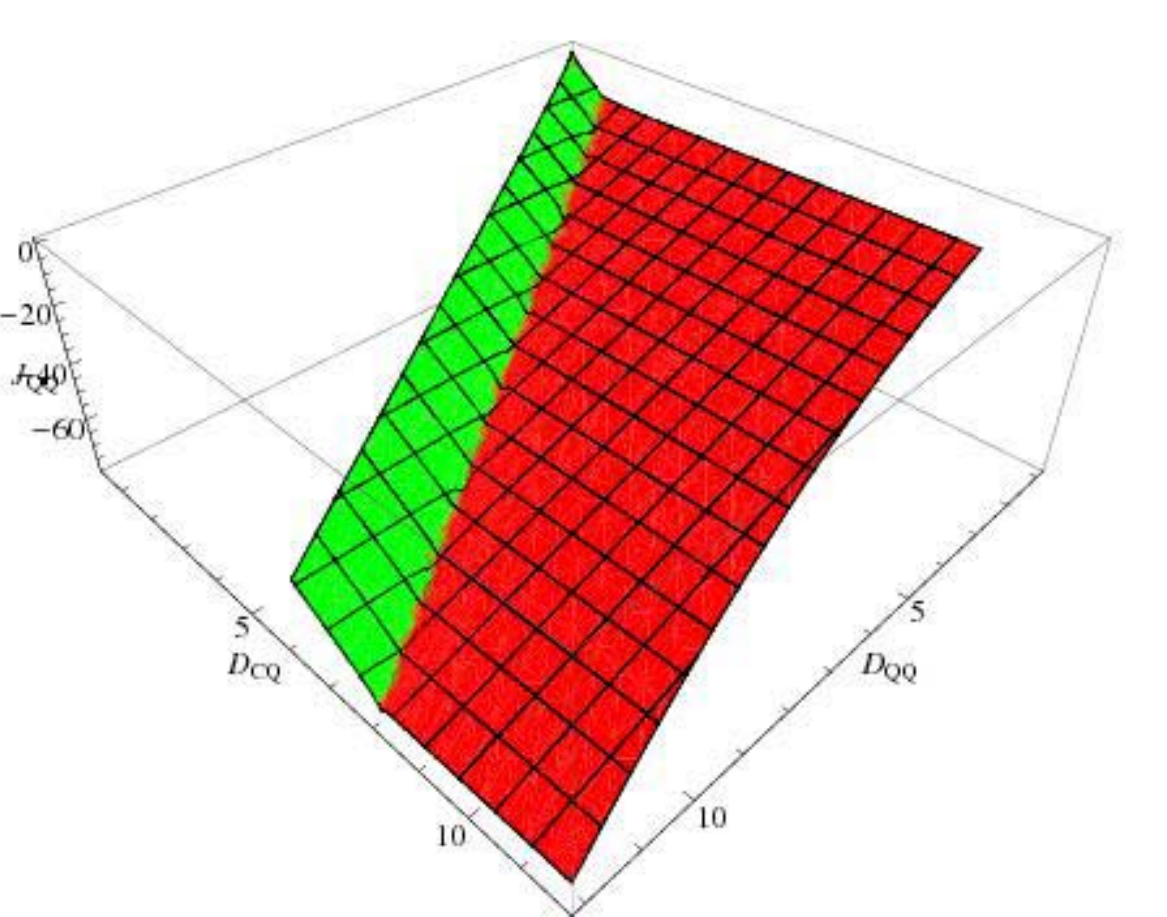}\\
(b)\\
\end{tabular}
\caption{(Colour online.) The exchange couplings (a) $J_{CQ}$ and (b) $J_{QQ}$ calculated by
perturbation theory as a function of qubit distance $D_{QQ}$ and qubit-control distance $D_{CQ}$. The sign of exchange coupling is coded in color; red: positive (antiferromagnetic) and green: negative
(ferromagnetic). $D_{QQ}$ varies from $1.0$ to $12.0$; $H$
varies from $0$ to $10$. }\label{pic:jcqjqq}
\end{figure}

For definiteness we take the following values for the parameters:
$U_{C}=1.0$, $U_{Q}=1.5$, $V_{QQ}=1/(2D_{QQ})$, $V_{CQ}=1/(2D_{CQ})$, $t_{QQ}=-10^{-2}e^{-2/\nu_gD_{QQ}}$,
$t_{CQ}=-e^{-(1/\nu_g+1/\nu_c)D_{CQ}}$, $\nu_g=0.7$, $\nu_c=1.0$. Using these parameters, we calculate
$J_{CQ},J_{QQ}$ shown in Figure~(\ref{pic:jcqjqq}).

Figure~\ref{pic:jcqjqq} shows us that when the qubit-qubit transfer amplitude $t_{QQ}$ is much weaker than the control-qubit transfer $t_{CQ}$, ring-exchange will dominate over simple two-centre exchange and hence lead to ferromagnetic exchange couplings. The variations of exchange coupling shown in Figure~\ref{pic:jcqjqq} are qualitatively similar to those shown in Figure~\ref{fig:JCQfig}(a) and Figure~\ref{fig:JQQfig}(a): when $D_{CQ}>D_{QQ}$, the exchange coupling is antiferromagnetic, while for some configurations where $D_{CQ}<<D_{QQ}$ the exchange coupling is ferromagnetic. Note that in the values assumed for the parameters, the prefactor for $t_{QQ}$ is chosen so that it is always much smaller than $t_{CQ}$; we find that the magnitude of this prefactor tunes the boundary between the ferromagnetic and antiferromagnetic regions, as might be expected because of its effect on the relative magnitudes of third- and fourth-order processes. For the excited control, $t_{CQ}$ is larger and the ferromagnetic region is consequently further extended.   Ultimately with these parameters a negative (ferromagnetic) part in the qubit-qubit exchange coupling $J_{QQ}$ arises at intermediate qubit-control separations.

\section{Statistical distribution of exchange interactions}\label{sec:statistics}
Since there is a strong configuration dependence of the exchange interactions it is natural to ask about their probability distributions at given impurity concentrations, since these distributions will determine the system's response to macroscopic, spatially averaging, probes.  We proceed as follows: we define a cluster of three spins to consist of a qubit atom, its nearest-neighbour qubit atom, and the control atom nearest to their mid-point.  Then, assuming the qubits and controls are independently and uniformly distributed (i.e. neglecting both defect-defect correlations and the unerlying lattice structure), the probability $P(D_{QQ},H)\,d D_{QQ}\,d H$ that the qubit-qubit distance is between $D_{QQ}$ and $D_{QQ}+\,d D_{QQ}$ while the control's distance from the mid-point is between $H$ and $H+\,d H$, is given by a simple generalization of the argument of Chandrasekhar \cite{chandra} as
\begin{equation}
P(D_{QQ},H)=16\pi^2D_{QQ}^2H^2n_Qn_C \mathrm{e}^{\left[-\frac{4\pi}{3}(n_QD_{QQ}^3+n_CH^3)\right]},
\end{equation}
where $n_Q$ and $n_C$ are the volume densities of qubit and control atoms respectively.  For purely exponential interactions between defect pairs, the averages over this type of distribution may be performed analytically \cite{stoneham83}; since we know that in this case the interactions depend on both $D_{QQ}$ and $H$, and are not purely exponential, we must resort to a numerical average. 

\begin{figure}[htbp]
\begin{tabular}{c}
\includegraphics[height=5cm]{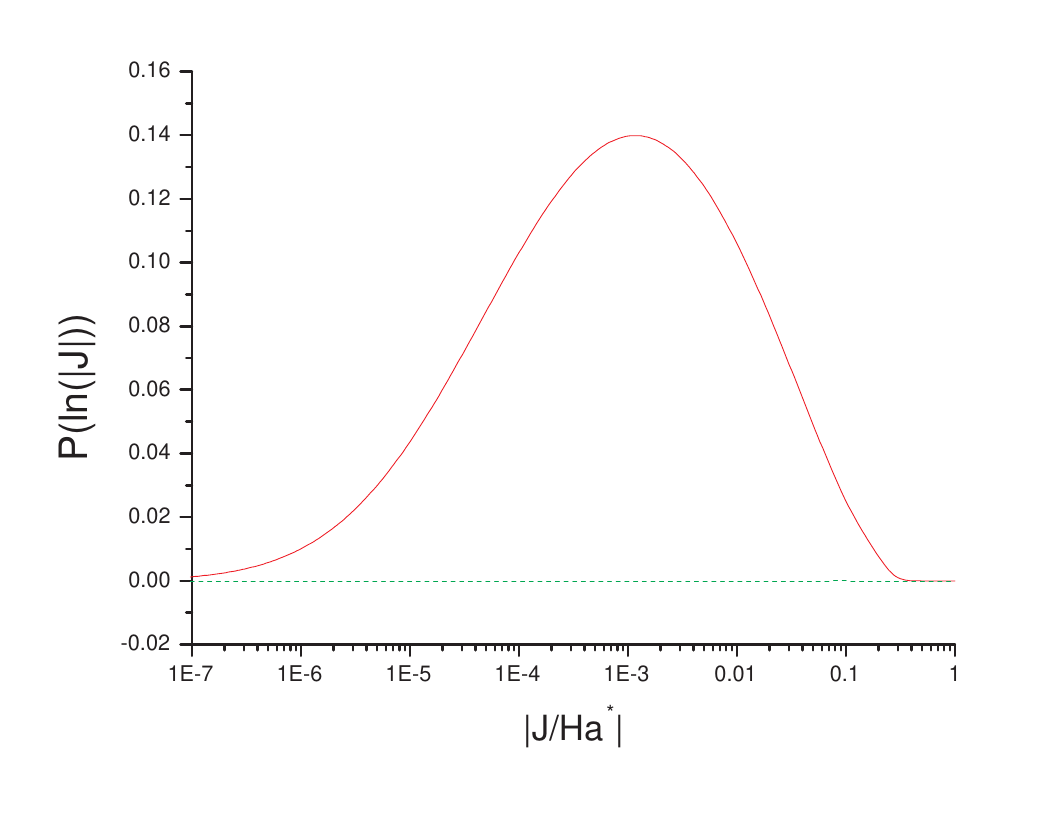}\\
(a)\\
\includegraphics[height=5cm]{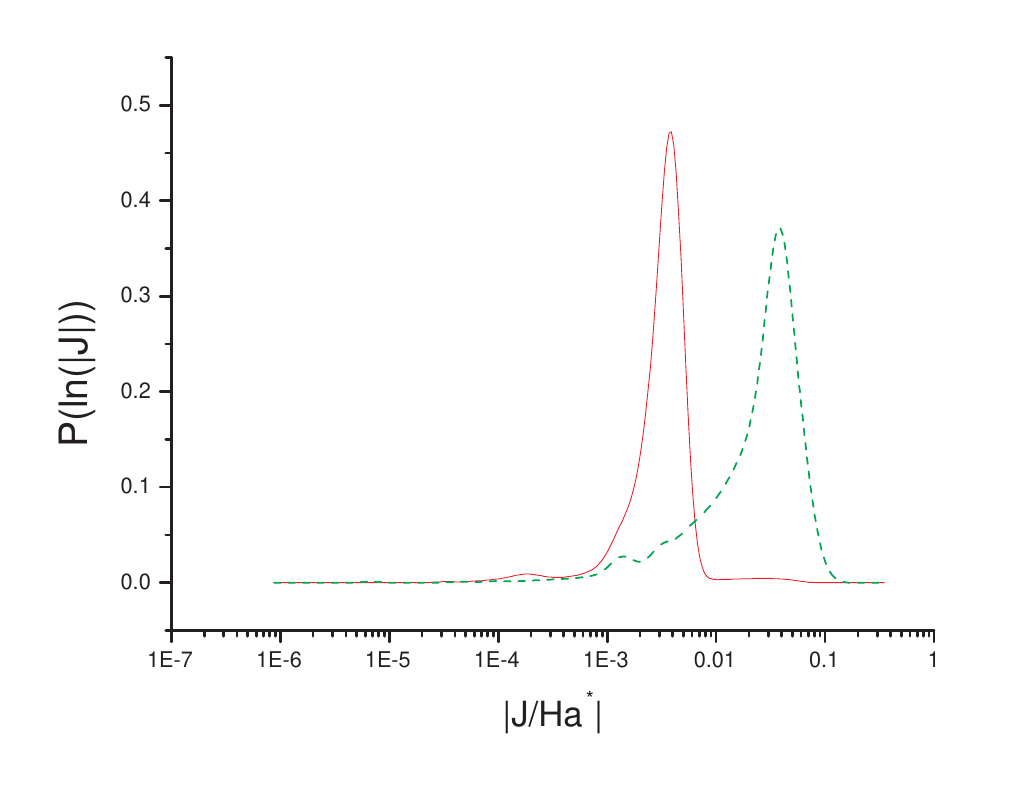}\\
(b)\\
\end{tabular}
\caption{(Colour online.) The probability distribution of $\log|J_{CQ}|$ for the control-qubit coupling, shown as a function of $J_{CQ}$ (on a logarithmic scale) for the control atom in (a) its ground state and (b) its excited state. Red solid curve: distribution of antiferromagnetic couplings; green dashed curve: ferromagnetic couplings.  The density of qubit atoms was taken as $0.00396\,(a_0^*)^{-3} =
5.42\times 10^{17}\,\mathrm{cm}^{-3}$, and the density of control atoms as $0.00198\,(a_0^*)^{-3} =
2.71\times 10^{17}\,\mathrm{cm}^{-3}$.  The plot was constructed by sampling over the $111\times101$-point grid used to create Figure~\ref{fig:JCQfig}, smoothed by convolving each point with a Gaussian in $\log|J_{CQ}|$ of standard deviation 0.224.}\label{pic:pjcq}
\end{figure}

\begin{figure}[htbp]
\begin{tabular}{c}
\includegraphics[height=5cm]{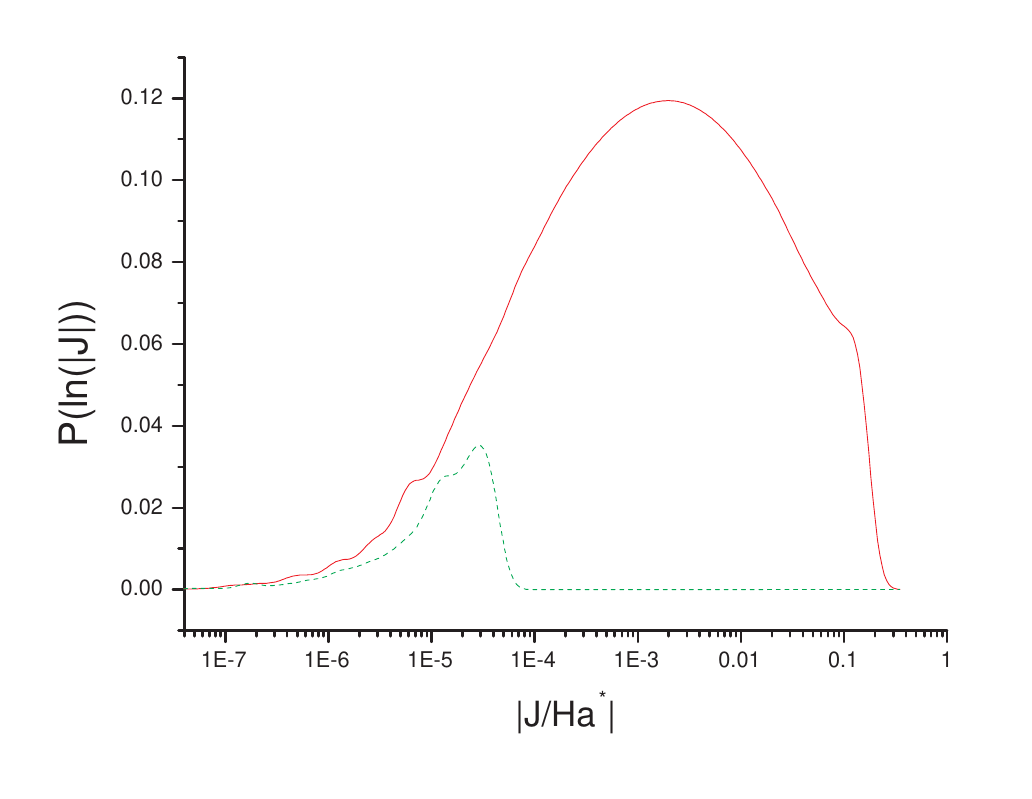}\\
(a)\\
\includegraphics[height=5cm]{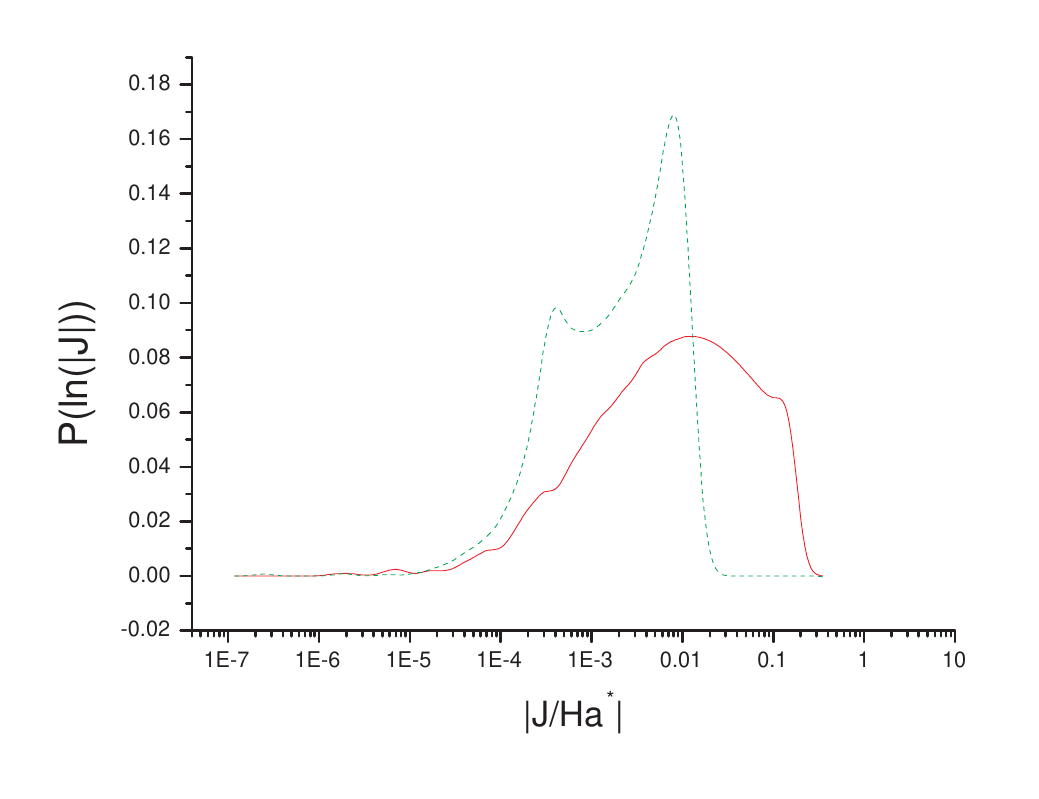}\\
(b)\\
\end{tabular}
\caption{(Colour online.) The probability distribution of $\log|J_{QQ}|$ for the qubit-qubit coupling as a function of $J_{QQ}$ (on logarithmic scale) for the control atom in (a) the ground state and (b) the excited state. Red solid curve: distribution of antiferromagnetic couplings; green dashed curve: ferromagnetic coupling distribution. The parameters used to construct the plot were as for Figure~\ref{pic:pjcq}.}\label{pic:pjqq}
\end{figure}

Our key approximation is to assume that the exchange couplings computed in \S\ref{sec:results} (i.e., in an isosceles configuration with both qubit-control distances equal) are representative of all configurations with the same values of $D_{QQ}$ and $H$; we suspect that this approximation will lead to excessive delocalization of the control electron and hence to an over-estimate of the exchange.   We compute the induced distribution of exchange couplings from our previous results by summing over the same mesh used to generate Figures~\ref{fig:JCQfig} and \ref{fig:JQQfig}, accumulating a suitably weighted histogram of exchange values by convoluting the delta-function contributions from each calculated point with a Guassian whose width is chosen to produce a smooth curve without introducing excessive broadening.  At the densities chosen, this mesh accounts for approximately 98\%\ of the normalization of $P(D_{QQ},H)$.  The results are shown in Figures~(\ref{pic:pjcq}) and (\ref{pic:pjqq}), in the form of distribution functions for $\log|J|$ with the contributions from positive and negative $J$ (i.e. antiferromagnetic and feroommagnetic parts) displayed separately. From Figure~\ref{pic:pjcq}(a), we can see that when the control is in the ground state the distribution of qubit-control exchange is overwhelmingly antiferromagnetic and qualitatively similar to those calculated for defect pairs of similar binding energies at comparable densities \cite{ourpaper1}. However, when the control is excited (Figure~\ref{pic:pjcq}(b)) the antiferromagnetic distribution narrows dramatically and shifts to higher $|J|$, both because of the slower decay with distance and because of the significant ferromagnetic region discussed in \S\ref{sec:results}.  At the same time the ferromagnetic part to the distribution grows.

Similarly, when the control is in the ground state the distribution of $J_{QQ}$ (Figure~\ref{pic:pjqq}) is predominantly antiferromagnetic and very close to that for deep-donor pairs at the same density \cite{ourpaper1}. The small ferromagnetic part is due to the ring exchange described in \S\ref{sec:ringexchange}. When the control is excited, this ferromagnetic region is greatly enlarged owing to the enhanced $t_{CQ}$, which leads to ring exchange dominating the two-donor exchange as illustrated in \S\ref{sec:ringexchange}.

\section{Discussion and conclusion}
We have described a variational method for calculating
the exchange interactions between three electrons in a control-qubit
system of the type envisaged for optically-controlled quantum information processing in silicon \cite{ams}. We have deliberately kept the model simple, yet rich enough to capture the coupling of the changing electron distribution to the spin configuration as the geometry and degree of excitation are altered.

The principal conclusion is that a three-centre system such as this is significantly different from the sum of two-electron exchange interactions that would arise from considering each defect pair separately.  In particular the excitation of the control electron can alter the `qubit-qubit` coupling $J_{QQ}$ as well as the `control-qubit' coupling $J_{CQ}$.  This is partly because of the delocalization of the electronic states across all three centres (particularly when the control electron is in an excited state), and partly because of the presence of three-centre ring-exchange processes (which we have also described perturbatively).  Most strikingly, we predict that either or both the control-qubit coupling $J_{CQ}$ and the qubit-qubit coupling $J_{QQ}$ can become ferromagnetic; this can happen through either of the above mechanisms. The excitation of the control greatly enhances both the ferromagnetism and the degree of charge transfer; however, it is interesting that $J_{QQ}$ can be ferromagnetic even when the control is in the ground state. The qualitative agreement between our variational results and those of the simple perturbation theory (which includes only the ring-exchange mechanism) shows that the latter is especially important when the control is in the ground state.  The ferromagnetic couplings we predict could be exploited for the initialization or manipulation of donor spins.

As might be expected, when the control electron moves further away, $J_{QQ}$ returns to the value expected for a deep donor pair without any control species \cite{ourpaper1}. This is also reflected in the statistical distributions of exchange values, where the distribution of $J_{QQ}$ with the control in the ground state is quite similar to the exchange distribution for deep-donor pairs calculated in \cite{ourpaper1}. However, when the control electron is located between the two qubits the magnitude of $J_{QQ}$ increases significantly, owing to the mediation by the control electron; this is especially significant when the qubit-qubit distance is large. In this limit $J_{QQ}$ can also become ferromagnetic, because the fourth-order terms containing $t_{CQ}$ which give a ferromagnetic contribution dominate. When the qubit-qubit distance is relatively small, the second-order terms involving $t_{QQ}$ dominate $J_{QQ}$, and the control electron is a minor perturbation on the two-qubit sub-system.

Our results show that the process of excitation and de-excitation can produce substantial but quite complex changes in the exchange couplings and charge distributions in these model defect complexes. They provide a theoretical demonstration of the switching of exchange couplings by excitation; a quantitative understanding of this process will have implications both for optically-controlled magnetism in silicon, and for quantum computation.

\begin{acknowledgments}
We wish to acknowledge the support of the UK Research Councils
Basic Technology Programme under grant GR/S23506.  We thank Gabriel Aeppli, Tony
Harker, Andy Kerridge, Chiranjib Mitra, Marshall Stoneham, and Dan Wheatley for
helpful discussions.

\end{acknowledgments}

\appendix
\section{Computing the coefficients in the Spin Hamiltonian}\label{app:extractH}
Equation~(\ref{eq:exchangeH}) is the most general
rotationally-invariant three-spin Hamiltonian that is even under
time reversal. Total spin angular momentum is a good quantum
number, and standard angular momentum coupling schemes
\cite{edmondsbook} show that three spin-1/2 particles can be
coupled to a quartet $|Q\rangle=\left\{\left
[\frac{1}{2},\frac{1}{2}\right
]_1,\frac{1}{2}\right\}_{\frac{3}{2}}$, and to two different
doublets, $|D_L\rangle=\left\{\left [\frac{1}{2},\frac{1}{2}\right
]_L,\frac{1}{2}\right\}_{\frac{1}{2}}$ distinguished by the
intermediate angular momentum $L$, which can be 0 or 1.

Since the full spin Hamiltonian (\ref{eq:exchangeH}) is traceless,
we may conveniently write its eigenvalues as
\begin{eqnarray}
E_{D_0}&=&-\epsilon-\delta/2\qquad\hbox{(doublet)}\nonumber\\
E_{D_1}&=&-\epsilon+\delta/2\qquad\hbox{(doublet)}\nonumber\\
E_{Q}&=&\epsilon\qquad\hbox{(quartet)}
\end{eqnarray}
It is sufficient to work with the three states having total spin component 1/2 ($|001\rangle$,
$|010\rangle$ and $|100\rangle$); they are only coupled amongst
themselves, and within this subspace the Hamiltonian is
\begin{widetext}
\begin{equation}
\frac{1}{4}\left [
\begin{array}{ccc}
J_{CQB}-J_{CQA}-J_{QQ} & 2J_{QQ} & 2J_{CQA}\\
2J_{QQ} & -J_{CQB}+J_{CQA}-J_{QQ} & 2J_{CQB}\\
2J_{CQA} & 2J_{CQB} & -J_{CQB}-J_{CQA}+J_{QQ}
\end{array}
\right]
\end{equation}
which has one eigenvalue
\begin{equation}
\frac{J_{QQ}+J_{CQA}+J_{CQB}}{4}=E_Q,
\end{equation}
(corresponding to the component of the quartet with magnetic
quantum number=1/2), while the other two eigenvalues are
\begin{equation}
-\epsilon\pm \frac{1}{2}\sqrt{(J_{CQB}^2+J_{CQA}^2+J_{QQ}^2)-J_{CQA}J_{QQ}-J_{QQ}J_{CQB}-J_{CQB}J_{CQA}}
\end{equation}
(corresponding to the two doublet states).
\end{widetext}

Now, in the symmetric case $J_{CQA}=J_{CQB}=J_{CQ}$, it is straightforward to show
that
\begin{equation}
{\cal L}=\langle(s_A+s_B)^2\rangle
\end{equation}
commutes with the Hamiltonian. Since ${\cal L}$ is the operator
which corresponds to the intermediate angular momentum $L$
described above, we may characterize the two doublet states by
their intermediate $L$ value. Once more it is convenient to work
within the spin 1/2 subspace, where a short calculation shows that
the doublet state with $L=0$ has the eigenvector
$\{1,-1,0\}^T/\sqrt{2}$, with eigenvalue
\begin{equation}
E_{D_0}=-\frac{3J_{QQ}}{4},
\end{equation}
whereas the doublet characterized by $L=1$ has eigenvalue
\begin{equation}
E_{D_1}=\frac{J_{QQ}}{4}-J_{CQ}
\end{equation}
with eigenvector $\{1,1,-2\}^T/\sqrt{6}$. The quartet state, of
course has eigenvalue
\begin{equation}
E_Q=\frac{J_{CQ}}{2}+\frac{J_{QQ}}{4},
\end{equation}
and eigenvector $\{1,1,1\}^T/\sqrt{3}$. Thus, the simplest way of
determining the exchange parameters $J_{QQ}$ and $J_{CQ}$ is to
identify the doublet state with $L=0$ and the quartet state and
use the relations (\ref{eq:equationsforcouplings}) given in the
main text.

Notice it is important to identify the doublet states by their
wavefunction; their energies alone are not sufficient. The
eigen{\em values} of the Hamiltonian defined by the exchange
parameters $J_{CQ}$ and $J_{QQ}$ are identical to those of the Hamiltonian
obtained by the substitutions
\begin{eqnarray}
J_{CQ}&\to& \frac{1}{3}(J_{CQ}+2J_{QQ});\nonumber\\
J_{QQ}&\to& \frac{1}{3}(4J_{CQ}-J_{QQ}).
\end{eqnarray}
However, the eigen{\em vectors} of the doublets interchange their
properties under this substitution. This is related to the sign
ambiguity in the square root in equation A3, which makes the sign
of $\delta$ in A1 ambiguous.

\section{Analytic results from perturbation theory}\label{sec:fullptresults}

\begin{widetext}
The lowest-order (second-order) contributions to the exchange are
  \begin{eqnarray}
  J_{CQ}^{(2)}&=&-\frac{2 t_{CQ}^2}{-U_{Q}+2 V_{CQ}-V_{QQ}}-\frac{2 t_{CQ}^2}{V_{QQ}-U_{C}};
  \end{eqnarray}
  \begin{eqnarray}
  J_{QQ}^{(2)}&=&-\frac{4 t_{QQ}^2}{V_{QQ}-U_{Q}}.
  \end{eqnarray}
The third-order contributions are
\begin{eqnarray}
J_{CQ}^{(3)}&=&\frac{2 t_{CQ}^2 t_{QQ}}{(V_{QQ}-U_{C})^2}-\frac{2 t_{CQ}^2 t_{QQ}}{(-U_{Q}+2
   V_{CQ}-V_{QQ})^2};
\end{eqnarray}
\begin{eqnarray}
J_{QQ}^{(3)}&=&\frac{4 t_{QQ} t_{CQ}^2}{(-U_{Q}+2 V_{CQ}-V_{QQ}) (V_{QQ}-U_{Q})}-\frac{4 t_{QQ}
   t_{CQ}^2}{(V_{QQ}-U_{C}) (V_{QQ}-U_{Q})}
   \\&&\nonumber+\frac{2 t_{QQ} t_{CQ}^2}{(-U_{Q}+2
   V_{CQ}-V_{QQ})^2}-\frac{2 t_{QQ} t_{CQ}^2}{(V_{QQ}-U_{C})^2},
\end{eqnarray}
while the fourth-order terms are
\begin{eqnarray}
J_{CQ}^{(4)}&=&-\frac{4 t_{CQ}^4}{(-U_{Q}+2 V_{CQ}-V_{QQ}) (V_{QQ}-U_{C}) (V_{QQ}-U_{Q})}-\frac{2
   t_{CQ}^4}{(-U_{Q}+2 V_{CQ}-V_{QQ})^2 (V_{QQ}-U_{Q})}
   \\&&\nonumber-\frac{2 t_{CQ}^4}{(V_{QQ}-U_{C})^2
   (V_{QQ}-U_{Q})}-\frac{2 t_{QQ}^2 t_{CQ}^2}{(-U_{Q}+2 V_{CQ}-V_{QQ})^3}-\frac{2 t_{QQ}^2
   t_{CQ}^2}{(V_{QQ}-U_{C})^3};
\end{eqnarray}
\begin{eqnarray}
J_{QQ}^{(4)}&=&\frac{4 t_{CQ}^4}{(-U_{Q}+2 V_{CQ}-V_{QQ}) (V_{QQ}-U_{C}) (V_{QQ}-U_{Q})}-\frac{4 t_{QQ}^2
   t_{CQ}^2}{(-U_{Q}+2 V_{CQ}-V_{QQ})^2 (V_{QQ}-U_{Q})}\\&&\nonumber-\frac{4 t_{QQ}^2
   t_{CQ}^2}{(V_{QQ}-U_{C})^2 (V_{QQ}-U_{Q})}
   -\frac{4 t_{QQ}^2 t_{CQ}^2}{(-U_{Q}+2
   V_{CQ}-V_{QQ}) (V_{QQ}-U_{Q})^2}-\frac{4 t_{QQ}^2 t_{CQ}^2}{(V_{QQ}-U_{C})
   (V_{QQ}-U_{Q})^2}.
\end{eqnarray}

\end{widetext}

\end{document}